\documentclass[aps,letterpaper,11pt,prx,onecolumn,floatfix]{revtex4-2} 
\usepackage[left=1.0in,right=1.0in,bottom=1.0in,top=1.0in]{geometry}
\usepackage{graphicx} 
\usepackage{braket}
\usepackage{multirow}
\usepackage{amsmath}
\usepackage{hyperref}
\usepackage{booktabs}
\usepackage{footnote}
\usepackage[caption=false]{subfig}
\usepackage{float}
\usepackage{tablefootnote}
\usepackage{xcolor}
\usepackage{float}

\begin{document}
\title{Fokker-Planck equation governing the distribution of walkers in AFQMC}
\author{Alfred Li}
\email{ali2@caltech.edu}
\affiliation{Department of Physics, California Institute of Technology, Pasadena, CA 91125, USA}

\author{Ankit Mahajan}
\affiliation{Department of Chemistry, Columbia University, New York, NY 10027, USA}

\author{Sandeep Sharma}
\email{sanshar@gmail.com}
\affiliation{Division of Chemistry and Chemical Engineering, California Institute of Technology, Pasadena, California 91125, USA}
\affiliation{Marcus Center for Theoretical Chemistry, Pasadena CA 91125, USA}
\begin{abstract}
Auxiliary-field quantum Monte Carlo (AFQMC) is typically formulated as an open-ended random walk in an overcomplete space of Slater determinants, implemented through a Langevin equation. However, the explicit form of the underlying Fokker-Planck equation governing the walker population distribution has remained unknown. In this paper, we derive the Fokker-Planck equation for AFQMC and propose a novel numerical scheme to solve it. The solution of the Fokker-Planck equation reveals the wavefunction actually sampled by the AFQMC algorithm. Interestingly, we find that even when the exact ground state is used as a guiding wavefunction in constrained path AFQMC, contrary to the common assumption, the wavefunction sampled by AFQMC is not exact. Beyond clarifying several fundamental aspects of AFQMC, the availability of a Fokker-Planck equation formulation opens new avenues for systematically improving its accuracy, which we outline in this paper. 
\end{abstract}
\maketitle

\section{Introduction}
In recent years auxiliary-field quantum Monte Carlo (AFQMC) has emerged as a powerful technique for solving problems in chemistry, materials science and condensed matter physics \cite{Motta_2018,zhang201315,PhysRevLett.90.136401,PhysRevB.80.214116,Purwanto_2008,PhysRevLett.100.126404,Purwanto_2011,PhysRevLett.104.116402,PhysRevLett.78.4486,PhysRevB.57.11980,PhysRevX.10.031016,PhysRevB.88.125132,PhysRevA.72.053610,PhysRevA.86.053606,PhysRevB.103.115123,PhysRevX.5.041041,PhysRevB.102.041106,mahajan2025beyond}. It belongs to a class of methods known as projector Monte Carlo that relies on stochastically 
propagating the imaginary time Schr{\"o}dinger equation to project out all excited states from an initial state so that in the long time limit one is left with the ground state \cite{PhysRevLett.45.566,reynolds1982fixed,RevModPhys.73.33,PhysRevLett.65.3437}. The imaginary time propagation is performed using an open-ended branching random walk algorithm (BRW) where walkers consisting of a Slater determinant, weight pair are updated stochastically (more details are provided in section~\ref{sec:cpmc}). The wavefunction sampled by the AFQMC algorithm at any given imaginary time can be written, in a statistical sense, as $|\Psi_{AF}(\tau)\rangle = \sum_i w_i(\tau) |\phi_i(\tau)\rangle $, where each walker is specified by the $\{w_i,|\phi_i\rangle\}$ pair. The projected energy is obtained as \begin{align}
    E_{AF}(\tau) = \frac{\langle\Psi_T|H|\Psi_{AF}(\tau)\rangle}{\langle\Psi_T|\Psi_{AF}(\tau)\rangle} = \frac{\sum_i w_i(\tau) \langle\Psi_T|H|\phi_i(\tau)\rangle}{\sum_i w_i(\tau)\langle\Psi_T|\phi_i(\tau)\rangle}\label{eq:projE}
\end{align}
against a reasonable trial state $|\Psi_T\rangle$, such as the Hartree-Fock state. For sufficiently large $\tau$ the energy $E_{AF}(\tau)$ provides an unbiased estimate of the ground state energy of the system, provided that no further approximations are introduced. By averaging over many such time slices one can obtain the ground state energy to a desired level of statistical accuracy. 

\begin{figure}
    \centering
    \includegraphics[width=0.4\linewidth]{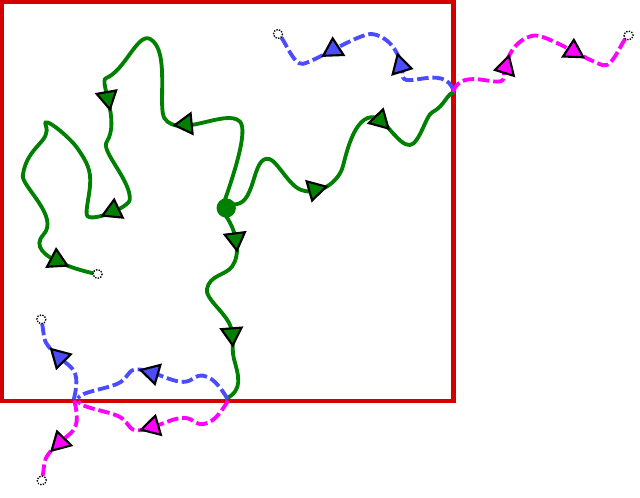}
    \caption{A 2-d representation of the random walk that is taken in the space of walkers. The arrows tracing out each path represents imaginary time or the number of steps in AFQMC. The thick red square represents the nodal surface of the ground state, that is, the set of walkers that are exactly orthogonal to the ground state. Three walkers, all starting from the same initial state are shown by green trajectories following stochastic paths. When some of the trajectories come in contact with the node the future propagation is shown with dotted lines. }
    \label{fig:nodal}
\end{figure}

Although this algorithm is exact in principle, in practice the BRW cannot be performed in an unbiased manner (except for systems with special symmetries, such as half-filled Hubbard model on the square lattice\cite{PhysRevB.31.4403}) because of the infamous sign problem \cite{PhysRevLett.67.3074,PhysRevLett.100.236404,PhysRevB.41.9301}, which prevents one from obtaining meaningful statistics. The origin of the sign problem can be understood by noticing that if $|\Psi_0\rangle$ is the ground state, then $-|\Psi_0\rangle$ is also the ground state and, unless additional constraints are introduced, the AFQMC algorithm will sample from a linear combination of these two states. As a result, the denominator of the energy estimator $\langle\Psi_T|\Psi_{AF}(\tau)\rangle$ will approach 0 with increasing $\tau$, causing a proliferation of the noise in the simulation. The sign problem can be eliminated if one could bias the walk towards positive $|\Psi_0\rangle$ and eliminate walkers from sampling regions where overlap with the ground state is negative. The practical manifestation of this idea is incorporated into the constrained-path approximation that eliminates all walkers during BRW that change sign relative to the trial state \cite{PhysRevLett.74.3652,PhysRevB.78.165101}. In other words if $\text{sgn}(\langle\Psi_T|\phi_i(\tau)\rangle) \neq  \text{sgn}(\langle\Psi_T|\phi_i(\tau+\delta\tau)\rangle)$ then the walker at time $\tau + \delta\tau$ given by $|\phi_i(\tau+\delta\tau)\rangle$ is eliminated. The constrained path approximation stabilizes the simulation at the cost of introducing an error in the ground state energy\cite{Zhang2004}.

The trial state imposes its sign structure on the AFQMC propagation and, therefore, plays a crucial role in determining its accuracy. In the limit when the trial state becomes equal to the ground state, the AFQMC algorithm yields the exact ground state energy. This follows from the projected energy expression in Eq. (\ref{eq:projE}), which reduces to the ground state energy irrespective of the walker and weight. This property is known as the \emph{zero-variance principle}\cite{assaraf1999zero} in quantum Monte Carlo. However, this observation alone does not establish that the AFQMC algorithm is sampling from the exact ground state. It is worth noting that a sister method known as diffusion Monte Carlo (DMC) is also a projector Monte Carlo method that uses a similar approach to avoid the sign problem, known as the \emph{fixed-node approximation}, to stabilize the BRW algorithm \cite{Annarelli_2024}. In the context of DMC, it is rigorously known that if the trial state is exact, then not only the energy but also the sampled wavefunction approaches the exact ground state. However, the argument is subtle and invokes the tiling theorem\cite{Ceperley1991}. Nevertheless, this line of reasoning often serves as a rationale for reducing the bias in AFQMC calculations by employing increasingly sophisticated trial states,\cite{PhysRevB.88.125132,shee2018phaseless,mahajan2022selected,jiang2024unbiasing} including those obtained from quantum computing algorithms.\cite{huggins2022unbiasing,zhao2025quantum}

In AFQMC, the standard justification provided is illustrated in Fig~\ref{fig:nodal}. Some of these trajectories show in the figure never come in contact with the node. When a walker comes in contact with the node, it is exactly orthogonal to the ground state. In exact propagation, such a walker contributes nothing to the ground state at all future imaginary times. However, in a stochastic setting it will contribute equally to $|\Psi_0\rangle$ and $-|\Psi_0\rangle$ (represented by the dotted magenta and blue curves) which manifests as noise. Thus if these walkers are removed from the simulation then that should introduce no additional bias in the simulation. This argument suggests that a constrained path simulation with the exact trial state should be unbiased. Nevertheless, as we demonstrate in this article, a residual bias in the AFQMC wavefunction remains. While the reasoning appears plausible, we will show that it in fact breaks down. Specifically, we will show that even when the exact trial state is used to impose the constraint, the AFQMC wavefunction can remain inexact and the associated error in the wavefunction can be substantial depending on the details of the Hamiltonian. Starting from the Langevin equation alone, it is difficult to determine if the bias appears because of numerical artifacts such as a finite time step, finite population, or if it is a fundamental shortcoming of the algorithm. To resolve this ambiguity, we derive a Fokker-Planck equation which, unlike the Langevin equation, can be solved deterministically using a finite-difference scheme. The systematic convergence of deterministic solutions with increasing discretization in the finite-difference scheme leaves little doubt that the bias is intrinsic to the method and further helps us understand its cause. 

The rest of the article is organized as follows. In Section~\ref{sec:cpmc}, we will outline the constrained-path AFQMC algorithm, which provides the necessary background for deriving the Fokker-Planck equation in Section~\ref{sec:fp}. This section also shows how the Fokker-Planck equation can be solved by using the finite-difference scheme. In Section~\ref{sec:results}, we solve the Fokker-Planck equation deterministically for small systems to demonstrate that its results agree with those from the AFQMC algorithm. Finally, Section~\ref{sec:conclusions} summarizes our conclusions and possible future directions.

\section{constrained-path AFQMC algorithm}\label{sec:cpmc}
In this article, we will focus entirely on the Hubbard model where the constrained-path AFQMC (cp-AFQMC) algorithm is employed. By contrast, in ab-initio Hamiltonians, one often uses the phaseless AFQMC (ph-AFQMC) algorithm \cite{PhysRevLett.90.136401,PhysRevB.88.125132}. cp-AFQMC is easier to analyze than ph-AFQMC, because the walkers are real-valued and the constraint is only invoked when a walker crosses the nodal boundary. The Hubbard model Hamiltonian can be written as a sum of one- and two-body terms:
\[ H = T_1 + T_2 = \sum_{ij \sigma } t_{ij} a_{i\sigma}^\dagger a_{j\sigma} + \sum_i U n_{i\alpha} n_{i\beta}\]
where $i, j$ are the site labels, $\sigma\in\{\alpha,\beta\}$ is the spin, $t_{ij}$ is the hopping term (here we allow long-range hopping), and $U$ is the on-site repulsion. The solution of the imaginary time Schr{\"o}dinger equation up to a normalization constant is formally given by
\begin{align}
|\Psi(\tau)\rangle =& \exp(-\tau (H-E_0)) |\Psi(0)\rangle    =\prod_{l=1}^{\tau/\delta\tau} \exp(-\delta\tau (H-E_0))|\Psi(0)\rangle \nonumber\\
\overset{(i)}{\approx}&\exp(\tau E_0)\prod_{l=1}^{\tau/\delta\tau}  \exp(-\delta\tau T_2)\exp(-\delta\tau T_1)|\Psi(0)\rangle \nonumber\\
\overset{(ii)}{=}&\exp(\tau (E_0 - UN/2)) \prod_{l=1}^{\tau/\delta\tau}  \exp(\delta\tau \sum_k H_k^2/2)\exp(-\delta\tau T_1)|\Psi(0)\rangle \nonumber\\
\overset{(iii)}{=}&\exp(\tau (E_0 - UN/2)) \prod_{l=1}^{\tau/\delta\tau} \left[\int \prod_k dx_k \frac{e^{-x_k^2/(2\delta\tau)}}{\sqrt{2\pi\delta\tau}} \exp(x_k H_k)\right]\exp(-\delta\tau T_1) |\Psi(0)\rangle \label{eq:HScont}
\end{align}
To obtain $(i)$ we use Trotter decomposition of the Hamiltonian with an associated error of order $O(\delta \tau^2) $ per time step 
For step $(ii)$ we use the identity $n_{i\alpha} n_{i\beta} = -\frac{1}{2}(n_{i\alpha} - n_{i\beta})^2 + \frac{1}{2}N_i$, where $N_i=n^2_{i\alpha} + n^2_{i\beta}=n_{i\alpha} + n_{i\beta}$, $N=\sum_i N_i$ and $H_k = \sqrt{U} (n_{k\alpha} - n_{k\beta})$. Finally, to obtain $(iii)$, we use the Hubbard-Stratonovich transformation\cite{1957SPhD....2..416S}, which does not incur additional Trotter errors here because all $H_k$ commute with each other. 

The overall algorithm proceeds as follows
\begin{enumerate}
    \item The walker $\{w_i(\tau), |\phi_i(\tau)\rangle\}$ at time slice $\tau$ is acted upon by the operator $\exp(-\delta\tau T_1)$ using the Thouless theorem \cite{THOULESS1960225}.
    \item A random number $x_k$ is sampled from the normal distribution with variance 1, and the operator $\exp(\sqrt{\delta\tau} \sum_k x_k H_k)$ is applied on the walker from the previous step. Again, this is possible to do due to the Thouless theorem.
    \item The single particle orbitals corresponding to the determinant $\exp(\sqrt{\delta\tau} \sum_k x_k H_k)\exp(-\delta\tau T_1)|\phi_i(\tau)\rangle $ are orthogonalized using QR-decomposition such that all the diagonal elements of the R-matrix are positive. The Q-matrix represents the normalized Slater determinant $|\phi_i(\tau+\delta\tau)\rangle $ and the product of the diagonal elements of the R-matrix along with the factor $\exp(\delta \tau (E_0 - UN/2))$ is multiplied to the weight $w_i(\tau)$ to obtain the weight $w_i(\tau+\delta\tau)$.
    \item These three steps can be performed sequentially multiple times to propagate the wavefunction forward to obtain unbiased results. However, as mentioned earlier, this leads to the sign problem. To control it, we simply set the weights $w_i(\tau+\delta\tau)$ equal to 0, if the overlap of $|\phi_i(\tau)\rangle$ and $|\phi_i(\tau+\delta\tau)\rangle$ with the trial state has a different sign. This is known as the \emph{constrained path} approximation. 
\end{enumerate}
An additional step called stochastic reconfiguration (SR) is  
performed to reduce weight fluctuations. In SR, the weights of all the walkers are kept roughly equal by stochastically eliminating walkers with a small weight and replicating walkers with a large weight \cite{PhysRevB.57.11446}. This step does not introduce additional bias. Finally, the value of $E_0$, which equals the ground state energy, is not known a priori and is dynamically updated so that the $L_1$ norm of all the total weight of walkers remains roughly constant. This step does introduce a bias, known as population control bias, which can be systematically made zero by increasing the number of walkers \cite{PhysRevLett.74.3652}.  

In practical calculations, another step called importance sampling is introduced that is meant to reduce the variance of the simulation \cite{PhysRevLett.74.3652}, however, it does not change the final results. For the purposes of this work, we will not apply importance sampling in any of the calculations to simplify the ensuing analysis. The result of our work also raises questions about the ability of importance sampling to reduce the variance in the calculation, which we discuss in more detail in Section~\ref{sec:conclusions}.


\section{Derivation and solution of Fokker-Planck equation for AFQMC}\label{sec:fp}
The algorithm outlined in the previous section can be viewed as a Langevin equation with an additional birth/death term. The walkers are distributed according to an underlying distribution which itself evolves according to an equation analogous to the Fokker-Planck equation \cite{PhysRevB.43.765}. In this section, we will derive this equation and then explicitly solve it for low-dimensional problems to obtain the wavefunction sampled by AFQMC. 

\subsection{Fokker-Planck equation for AFQMC}
We first note that the wavefunction sampled by AFQMC at any imaginary time $\tau$ can be written as 
\begin{equation}
    \ket{\psi(\tau)} = \int\psi(\vec{\theta},\tau)\ket{\phi(\vec{\theta})}d\vec{\theta} \label{eq:state}
\end{equation}
where $\ket{\phi(\vec{\theta})}$ denotes a normalized Slater determinant on the Grassmann manifold and $\vec{\theta}$ is used to parametrize this space~\cite{Motta_2018}. Note that these determinants form an overcomplete basis set for representing the wavefunction. In section~\ref{sec:parametrization}, we will show how Givens rotations can be used as the parameters, but for now, we keep the discussion general. The goal is to obtain a Fokker-Planck equation, a partial differential equation that governs the evolution of $\psi(\vec{\theta},\tau)$  as a function of parameters and imaginary time. To derive this equation, we analyze how $\psi(\vec{\theta},\tau)$ is updated under the action of one-body and two-body parts of the propagator. Before moving to the general case, we will show how the state of a system containing 1 electron in 2 orbitals gets updated due to a one-body operator. Here, we can derive analytic results and build some intuition about what the various terms mean. 
\subsubsection{Action of one-body propagator on one-electron in two-orbitals}
A normalized single Slater determinant can be written as
\[|\phi(\theta)\rangle=\left[\begin{array}{c}
\cos(\theta)\\
\sin(\theta)
\end{array}\right]\]
and by varying $\theta$ between $-\pi/2$ to $\pi/2$ we can represent any normalized determinant.
For ease of exposition, let us assume the one-body operator to be 
\[T=\left[\begin{array}{cc}
a & b\\
b & a
\end{array}\right]\]
Then the action of the propagator on the state for a small time step is
\begin{align}
e^{\delta\tau T}|\phi(\theta)\rangle&=(1+\delta\tau T)|\phi(\theta)\rangle\nonumber\\
&=\left[\begin{array}{c}
\cos(\theta)\\
\sin(\theta)
\end{array}\right]+\delta\tau\left[\begin{array}{c}
a\cos(\theta)+b\sin(\theta)\\
b\cos(\theta)+a\sin(\theta)
\end{array}\right]\nonumber\\
&=(1+a\delta\tau+b\delta\tau\sin(2\theta))\left[\begin{array}{c}
\cos(\theta)\\
\sin(\theta)
\end{array}\right]-b\delta\tau\cos(2\theta)\left[\begin{array}{c}
\sin(\theta)\\
-\cos(\theta)
\end{array}\right]    \label{eq:update1}
\end{align}

This is a linear combination of two states (the original state and a single excitation orthogonal to it). There are an infinite number of ways in which this state can be written as a linear combination of states in the Grassmann manifold because of its overcompleteness. However, the AFQMC algorithm writes this wavefunction in a very specific way, and that is given by (one can confirm that the expression below is equal to the one above up to linear order in $\delta\tau$):
\begin{align}
e^{\delta\tau T}|\phi(\theta)\rangle&=(1+a\delta\tau+b\delta\tau\sin(2\theta))\left[\begin{array}{c}
\cos(\theta+\delta\tau b\cos(2\theta))\\
\sin(\theta+\delta\tau b\cos(2\theta))
\end{array}\right]\nonumber\\
&=(1+w(\theta)\delta\tau)|\phi(\theta+a(\theta)\delta\tau)\rangle    
\end{align}
where $w(\theta)=a+b\sin(2\theta),~a(\theta)=b\cos(2\theta)$ are respectively functions of the operator $\theta$  and specify the first order change in the weight and normalized Slater determinant due to the action of the operator. 

The full wavefunction is given by (\ref{eq:state}) and we will act the one-body operator on it:
\begin{align}
    e^{\delta\tau T}|\psi(\tau)\rangle&=\int\psi(\theta,\tau)e^{\delta\tau T}|\phi(\theta)\rangle d\theta\nonumber\\
&=\int\psi(\theta,\tau)\left(1+w(\theta)\delta\tau\right)|\phi(\theta+a(\theta)\delta\tau)\rangle d\theta\label{eq:first1e2o}
\end{align}
To make progress, let us introduce a new variable
\begin{align}
    \theta'&=\theta+a(\theta)\delta\tau \label{eq:changeVar}
\end{align}
In what follows we will have to invert this equation up to $O(\delta\tau)$. We will do this using perturbation theory by proposing \[\theta=\theta'+\delta\tau f(\theta')\]
plugging it back into (\ref{eq:changeVar}) we obtain
\begin{align}
    \theta'&=\theta'+\delta\tau f(\theta')+a(\theta'+\delta\tau f(\theta'))\delta\tau\nonumber\\
    &=\theta'+\delta\tau(f(\theta')+a(\theta'))+O(\delta\tau^{2})
\end{align} 
Equating term of $O(\delta\tau)$ equal to zero implies that $f(\theta)=-a(\theta)$. Thus we get $\theta'=\theta+a(\theta)\delta\tau$  and also $\theta=\theta'-\delta\tau a(\theta')$ which tells us 

\begin{align}
d\theta=(1-\delta\tau a'(\theta'))d\theta'\label{eq:jacobian}
\end{align}

Now performing a change of variables in (\ref{eq:first1e2o}) we get:
\begin{align}
e^{\delta\tau T}|\psi(\tau)\rangle&=\int\psi(\theta'-a(\theta')\delta\tau,\tau)\left(1+w(\theta'-\delta\tau a(\theta'))\delta\tau\right)|\phi(\theta')\rangle(1-\delta\tau a'(\theta'))d\theta'\nonumber\\
&=\int\left[\psi(\theta,\tau)-\frac{\partial\psi}{\partial\theta}a(\theta)\delta\tau\right]\left[1+ w(\theta)\delta\tau\right](1-\delta\tau a'(\theta))|\phi(\theta)\rangle d\theta   \label{eq:2a} 
\end{align}
In the second equation we have just relabeled the variable $\theta'$ to $\theta$. Another way to write the updated wavefunction is:
\begin{align}
e^{\delta\tau T}|\psi(\tau)\rangle=\int\left(\psi(\theta,\tau)+\frac{\partial\psi}{\partial\tau}\delta\tau\right)|\phi(\theta)\rangle d\theta \label{eq:2b} 
\end{align}
Let us equate terms in (\ref{eq:2a}) and (\ref{eq:2b}) which gives us
\begin{align}
\frac{\partial\psi}{\partial\tau}=&-a(\theta)\frac{\partial(\psi)}{\partial\theta}+(w(\theta)-a'(\theta))\psi\nonumber\\
\frac{\partial\psi}{\partial\tau}=&-\frac{\partial(a(\theta)\psi)}{\partial\theta}+w(\theta)\psi\nonumber\\
\frac{\partial\psi}{\partial\tau}=&-\frac{\partial(b\psi\cos(2\theta))}{\partial\theta}+(a+b\sin(2\theta))\psi \label{eq:simple1e}
\end{align}

This equation is the Fokker-Planck equation for the evolution of $\psi(\theta,\tau)$ as a function of $\tau$ due to the action of a one-body operator. Note that we were able to obtain the functions $a(\theta),w(\theta)$ in a closed form for this case. For more general cases this is not possible and we will use algorithmic differentiation to obtain them as will be explained in Section~\ref{sec:parametrization}. Next we obtain the equation for a more general many-electron system.

\subsubsection{Action of one-body propagator}
First let us look at the action of the one-body operator on a single Slater determinant:
\begin{align}
    e^{\delta \tau T_1}|\phi(\vec{\theta})\rangle = (1+w(\vec{\theta})\delta\tau)|\phi(\vec{\theta} + \vec{\chi}(\vec{\theta})\delta\tau)\rangle 
\end{align}
where $\vec{\theta} + \vec{\chi}(\vec{\theta})\delta\tau$ represents the parameters corresponding to an updated normalized determinant and $(1+w(\vec{\theta})\delta\tau)$ represents the norm of the determinant after time $\delta\tau$. The change in norm results from the fact that the propagator is not a unitary operator. In Section~\ref{sec:parametrization} we will describe how the values of $w(\vec{\theta})$ and $\vec{\chi}(\vec{\theta})$ can be obtained. 

Consequently, the action of the one-body propagator on the total wavefunction $\ket{\psi(\tau)}$ over a time $\delta\tau$ is, to first order,

\begin{align}
    e^{\delta\tau T_{1}}\ket{\psi(\tau)} =& \int\psi(\vec{\theta},\tau)e^{\delta\tau T_{1}}\ket{\phi(\vec{\theta})}d\vec{\theta} = \int\psi(\vec{\theta},\tau) (1+w(\vec{\theta})\delta\tau)|\phi(\vec{\theta} + \vec{\chi}(\vec{\theta})\delta\tau)\rangle  d\vec{\theta}\label{eq:start}
\end{align}
Before proceeding further we will introduce a new variable
\begin{align}
    \vec{\theta}' = \vec{\theta} + \vec{\chi}(\vec{\theta})\delta \tau\label{eq:forward}
\end{align}
and the inverse of this equation is
\begin{align}
    \vec{\theta} = \vec{\theta}' - \vec{\chi}(\vec{\theta}')\delta \tau \label{eq:inv}
\end{align}
which can be shown by substituting the value of $\vec{\theta}$ from (\ref{eq:inv}) into (\ref{eq:forward}) and Taylor expanding $\vec{\chi}$ up to linear order in $\delta\tau$. Using the change of variable in (\ref{eq:start}) will give rise to a Jacobian term:
\begin{align}
    d\vec{\theta} =& \left|\mathcal{J}\right|d\vec{\theta}'\nonumber\\
    \mathcal{J}_{\alpha,\beta} =& \delta_{\alpha,\beta} - (\partial_\alpha \chi_\beta) \delta\tau \nonumber\\
    \left|\mathcal{J}\right| =& (1 - \delta\tau(\partial_\alpha \chi_\alpha)) 
\end{align}
where $\partial_\alpha\chi_\beta=\frac{\partial \chi_\beta(\vec{\theta})}{\partial \theta_\alpha}$, in the third line we evaluate the Jacobian correct up to linear order in $\delta\tau$ and repeated indices imply a summation. Using all these in (\ref{eq:start}) and replacing the label $\theta'$ with $\theta$ we obtain
\begin{align}
e^{\delta\tau T_{1}}\ket{\psi(\tau)} =& \int \psi\left(\vec{\theta} - \vec{\chi}(\vec{\theta})\delta \tau, \tau\right)   \left(1+w(\vec{\theta} - \vec{\chi}(\vec{\theta})\delta \tau)\delta\tau\right)(1 - \delta\tau(\partial_\alpha \chi_\alpha))|\phi(\vec{\theta})\rangle  d\vec{\theta}\nonumber\\
=&\int\left(\psi - (\partial_\alpha \psi) \chi_\alpha \delta \tau\right)\left(1+ w  \delta \tau\right)(1 - \delta\tau(\partial_\alpha \chi_\alpha))|\phi(\vec{\theta})\rangle  d\vec{\theta}\nonumber\\
=&\int \left[\psi + \left(- (\partial_\alpha \psi) \chi_\alpha + \psi (w- \partial_\alpha \chi_\alpha)\right)\delta\tau \right]|\phi(\vec{\theta})\rangle  d\vec{\theta}\label{eq:oneBody1}
\end{align}
where in the second line we have performed Taylor series expansion of $\psi$ and $w$ up to linear order, dropped the explicit dependence of $\psi,w,\chi$ on $\vec{\theta}$ and summation over repeated indices is implied. In the third line we have eliminated all terms that are higher than linear order in $\delta\tau$. 

An alternate way of writing the action is to update the field $\psi(\vec{\theta},\tau)$ 
\begin{equation}
    e^{\delta\tau T_{1}}\ket{\psi(\tau)} = (1+\delta\tau T_{1})\ket{\psi(\tau)} =  \int\left(\psi(\vec{\theta},\tau)+\delta\tau\frac{\partial\psi(\vec{\theta},\tau)}{\partial\tau}\right)\ket{\phi(\vec{\theta)}}d\vec{\theta}\label{eq:oneBody2}
\end{equation}
Comparing the integrands in (\ref{eq:oneBody1}) and (\ref{eq:oneBody2}) and equating the terms linear in $\delta\tau$ we obtain the partial differential equation:
\begin{equation}
\label{one-spin one-body PDE}
    \frac{\partial\psi}{\partial\tau} = -\sum_{\alpha}\frac{\partial (\psi\chi_\alpha)}{\partial \theta_{\alpha }} + w\psi 
\end{equation}

\subsubsection{Action of two-body propagator}
For the two-body term we again look at the action of the propagator (after performing the Hubbard-Stratonovich transformation) on a single Slater determinant $|\psi(\vec{\theta})\rangle$ up to second order in the auxiliary fields:
\begin{align}
    e^{x_kH_k}|\phi(\vec{\theta})\rangle=&\left(1+g_i(\vec{\theta})x_i+h_{ij}(\vec{\theta})x_ix_j\right)|\phi(\vec{\theta}+ \vec{\gamma}_i(\vec{\theta})x_i+\vec{\kappa}_{ij}(\vec{\theta})x_ix_j)\rangle
\end{align}
where $g,h,\vec{\gamma},\vec{\kappa}$ represent the first and second order changes in the weight and position of the Slater determinant after the action of the propagator and repeated indices are assumed to be summed. Terms up to second order in $x$ need to be kept to obtain results that are correct to linear order in $\delta\tau$, as will become clear when we integrate the auxiliary fields. In what follows we will use Roman subscripts for auxiliary fields and Greek subscripts for components of the parameter $\vec{\theta}$. 

The action of the propagator on the state can be written as
\begin{align}
 &e^{\frac{1}{2}\delta\tau\sum_k H_{k}^2} \ket{\psi(\tau)}\nonumber\\
\overset{(i)}{=}&\int \prod_k dx_{k} \frac{e^{-x_{k}^2/(2\delta\tau)}}{\sqrt{2\pi\delta\tau}}\int\psi(\vec{\theta},\tau)\left[e^{ x_{k}H_k}\ket{\phi(\vec{\theta})}\right]d\vec{\theta}\nonumber\\ 
=& \int \prod_k dx_{k} \frac{e^{-x_{k}^2/(2\delta\tau)}}{\sqrt{2\pi\delta\tau}}\int\psi(\vec{\theta},\tau)\left(1+g_i(\vec{\theta})x_i+h_{ij}(\vec{\theta})x_ix_j\right)|\phi(\vec{\theta}+ \vec{\gamma}_i(\vec{\theta})x_i+\vec{\kappa}_{ij}(\vec{\theta})x_ix_j)\rangle d\vec{\theta}\label{eq:start2} 
\end{align}
Let us introduce a new variable 
\begin{align}
    \theta_{\alpha}'=\theta_{\alpha}+\gamma_{\alpha i}x_{i}+\kappa_{\alpha ij}x_{i}x_{j}\label{eq:atob}
\end{align}
where, we have eliminated the explicit dependence of $\gamma,\kappa$  on $\theta$  and $x$. We can invert the equation which give us
\begin{align}
    \theta_{\alpha}&=\theta_{\alpha}'-\gamma_{\alpha i}x_{i}-\left(\kappa_{\alpha,ij}-(\partial_{\beta}\gamma_{\alpha i})\gamma_{\beta j}\right)x_{i}x_{j}\label{eq:btoa}
\end{align}
where, $\gamma_{\alpha i},\kappa_{\alpha,ij}$ are function of $\theta'$ on the right hand side. The correctness of this equation can be confirmed by substituting (\ref{eq:btoa}) into (\ref{eq:atob}) and doing Taylor series expansion of $\gamma$ to linear order in $x$. To simplify the notation in subsequent equations we will define 
\begin{align}
    \eta_{\alpha ij} =& \kappa_{\alpha,ij}-(\partial_{\beta}\gamma_{\alpha i})\gamma_{\beta j}
\end{align}
Doing this change of variables in (\ref{eq:start2}) will give rise to a Jacobian
\begin{align}
    d\vec{\theta}=&|\mathcal{J}| d\vec{\theta}'\nonumber\\
    \mathcal{J}_{\alpha\beta} =& \delta_{\alpha\beta} - (\partial_\alpha \gamma_{\beta i})x_i -(\partial_\alpha \eta_{\beta ij})x_i x_j\nonumber\\
    |\mathcal{J}| =& 1 - (\partial_\alpha \gamma_{\alpha i})x_i - (\partial_\alpha \eta_{\alpha ij})x_i x_j +\frac{1}{2} \left\{(\partial_\alpha \gamma_{\alpha i})(\partial_\beta \gamma_{\beta j})x_ix_j\right\} - \frac{1}{2}\left\{(\partial_\alpha \gamma_{\beta i}) (\partial_\beta \gamma_{\alpha j}) x_ix_j\right\}\label{eq:jacobian2O}
\end{align}
where we have used the identity $\det(I+A)=1 + \text{tr}A + \frac{1}{2}\left((\text{tr}A)^2 - \text{tr}A^2\right) + O(A^3)$ 

Using these results in (\ref{eq:start2}) we obtain
\begin{align}
& \int \prod_k dx_{k} \frac{e^{-x_{k}^2/(2\delta\tau)}}{\sqrt{2\pi\delta\tau}}\int\psi(\vec{\theta},\tau)\left(1+g_i(\vec{\theta})x_i+h_{ij}(\vec{\theta})x_ix_j\right)|\phi(\vec{\theta}+ \vec{\gamma}_i(\vec{\theta})x_i+\vec{\kappa}_{ij}(\vec{\theta})x_ix_j)\rangle d\vec{\theta}\nonumber\\
=&\int \prod_k dx_{k} \frac{e^{-x_{k}^2/(2\delta\tau)}}{\sqrt{2\pi\delta\tau}}\int\psi(\vec{\theta}-\vec{\gamma}_ix_i -\vec{\eta}_{ij}x_ix_j,\tau)\left(1+g_i(\vec{\theta}-\vec{\gamma}_ix_i)x_i+h_{ij}(\vec{\theta})x_ix_j\right)|\mathcal{J}(\vec{\theta},\vec{x})||\phi(\vec{\theta})\rangle d\vec{\theta}\nonumber\\
=&\int \prod_k dx_{k} \frac{e^{-x_{k}^2/(2\delta\tau)}}{\sqrt{2\pi\delta\tau}} \int\left(\psi(\vec{\theta})-(\partial_\alpha \psi) (\gamma_{\alpha i}x_i +\eta_{\alpha ij}x_ix_j)+\frac{1}{2}(\partial_\alpha\partial_\beta\psi)\gamma_{\alpha i}\gamma_{\beta j}x_ix_j \right)\nonumber\\
&\left(1+g_ix_i-(\partial_\alpha g_i) \gamma_{\alpha j}x_ix_j+h_{ij}x_ix_j\right)|\mathcal{J}(\vec{\theta},\vec{x})||\phi(\vec{\theta})\rangle d\vec{\theta}\nonumber\\
\overset{(i)}{=}&\int\left(\psi(\vec{\theta})+\left[-(\partial_\alpha \psi) (\gamma_{\alpha i}g_i-\gamma_{\alpha i}(\partial_\beta \gamma_{\beta i}) + \eta_{\alpha ii})+\frac{1}{2}(\partial_\alpha\partial_\beta\psi)\gamma_{\alpha i}\gamma_{\beta i} \right. \right.\nonumber\\
&+ \left.\left. \psi \left(h_{ii} - g_i (\partial_\alpha \gamma_{\alpha i}) - (\partial_\alpha g_i)\gamma_{\alpha i}-\partial_\alpha\eta_{\alpha ii}+\frac{1}{2}(\partial_\alpha \gamma_{\alpha i})(\partial_\beta \gamma_{\beta i})-\frac{1}{2}(\partial_\alpha \gamma_{\beta i}\partial_\beta \gamma_{\alpha i})\right)\right]\delta\tau \right) |\phi(\vec{\theta})\rangle d\vec{\theta}\nonumber\\
\overset{(ii)}{=}&\int\left(\psi(\vec{\theta})+\left[-\partial_\alpha \left\{\psi (\gamma_{\alpha i}g_i + \kappa_{\alpha ii})\right\}+\frac{1}{2}\partial_\alpha\partial_\beta\left\{\psi \gamma_{\alpha i}\gamma_{\beta i}\right\} +h_{ii}\psi\right]\delta\tau \right)d\vec{\theta}\label{eq:twoBody1}
\end{align}
In step $(i)$ we have performed a Gaussian integrals and only terms that are quadratic in $x$ with repeating indices remain. The detailed derivation of step $(ii)$ is given in Appendix~\ref{eq:app1}. 

The change in wavefunction can also be written as follows
\begin{equation}
    e^{-\delta\tau T_2}\ket{\psi(\tau)} = (1-\delta\tau T_2)\ket{\psi(\tau)} = \int \left(\psi(\vec{\theta},\tau)-\delta\tau\frac{\partial\psi(\vec{\theta},\tau)}{\partial\tau}\right)\ket{\phi(\vec{\theta})}d\vec{\theta}\label{eq:twoBody2}
\end{equation}
Comparing terms linear in $\delta\tau$ between (\ref{eq:twoBody1}) and (\ref{eq:twoBody2}) gives us the PDE. 

\begin{equation}
\label{one-spin two-body PDE}
    \frac{\partial\psi}{\partial\tau} = -\frac{1}{2}\sum_{\alpha\beta i}\frac{\partial^2(\psi\gamma_{\alpha i}\gamma_{\beta i})}{\partial \theta_{\alpha}\partial \theta_{\beta}}
    +\sum_{\alpha i}\frac{\partial\left(\psi(\gamma_{\alpha i}g_i + \kappa_{\alpha ii})\right)}{\partial \theta_{\alpha }}
    -\sum_i h_{ii}\psi
\end{equation}

\subsubsection{Fokker-Planck equation}

Taking the one-body and two-body terms together, we end up with the following equation for the evolution of the wavefunction:
\begin{align}
\label{one spin sector PDE}
    \frac{\partial\psi}{\partial\tau} = -\frac{1}{2}\sum_{\alpha\beta}\frac{\partial^2(\psi\sum_i \gamma_{\alpha i}\gamma_{\beta i})}{\partial \theta_{\alpha}\partial \theta_{\beta}}
    +\sum_{\alpha }\frac{\partial\left(\psi(-\chi_\alpha + \sum_i\gamma_{\alpha i}g_i + \kappa_{\alpha ii})\right)}{\partial \theta_{\alpha }}
    + (w-\sum_ih_{ii})\psi
\end{align}
For compactness, we introduce:
\begin{align}
\mathcal{D}_{\alpha\beta} =\sum_{i}\gamma_{\alpha i}\gamma_{\beta i},~~\mathcal{V}_\alpha=-\chi_\alpha + \sum_i\gamma_{\alpha i}g_i + \kappa_{\alpha ii},~~\mathcal{S}=w-\sum_ih_{ii}
\end{align}
Then the PDE can be written in a Fokker–Planck form with a birth/death term: 
\begin{align}
    \frac{\partial\psi}{\partial\tau}=& -\frac{1}{2}	\sum_{\alpha\beta} \frac{\partial^{2}(\psi\mathcal{D}_{\alpha\beta})}{\partial \theta_{\alpha}\partial \theta_{\beta}} + \sum_{\alpha} \frac{\partial(\psi\mathcal{V}_\alpha)}{\partial \theta_{\alpha}} + \mathcal{S}\psi\label{eq:fokker-planck}
\end{align}

Here, $\mathcal{D}_{\alpha\beta}$ is the position dependent anisotropic diffusion coefficient, $\mathcal{V}_{\alpha}$ is the position dependent anisotropic drift term and $\mathcal{S}$ is the source term. 

In the next section, we will introduce the Givens rotation as a concrete parametrization of $\vec{\theta}$, explain how the coefficients $\mathcal{D}_{\alpha\beta},\  \mathcal{S}_{\alpha},\ \mathcal{S}$ are obtained, and finally describe how the equation is solved using a finite-difference method. 

\subsection{Parameterizing Grassmann Manifold}\label{sec:parametrization}
The first step in solving the Fokker-Planck equation is to choose a parametrization for the {real Grassmannian manifold \cite{Chiumiento_2012}, the space of all possible unique Slater determinants (SD) in which the AFQMC algorithm is performing its random walk. 

A real SD with a unit norm can be parametrized by two matrices ($U^\alpha$ and $U^\beta$) with orthogonal columns of size $M \times N_\alpha$ and $M \times N_\beta$ respectively, where $M$ is the number of basis functions and $N_\alpha, N_\beta$ are the number of $\alpha$ and $\beta$ electrons in the system respectively. 
\begin{align}
    |\phi\rangle =& \prod_i^{N_\alpha} a_{i,\alpha}^\dagger \prod_j^{N_\beta} a_{j,\beta}^\dagger |0\rangle \\
    a_{i,\sigma}^\dagger =&\sum_p^M U_{p,i}^\sigma a_{p,\sigma}^\dagger
\end{align}
where $\sigma\in\{\alpha,\beta\}$, $U^\sigma$ is a matrix with orthogonal columns 
that transforms the creation operators $a_{p,\sigma}^\dagger$ corresponding to the underlying basis functions ($\vartheta_p$) into the creation operator $a_{i,\sigma}^\dagger$ corresponding to the orbital ($\varphi_{i\sigma}$) of the SD $|\phi\rangle$. In the following, for convenience we will drop the spin symbol $\sigma$ from all quantities.

The $MN$ elements of $U$ constitute a redundant parametrization of the SD: for example, a transformation among its columns does not change the determinant except for an overall sign. A non-redundant parametrization for all determinants not orthogonal to a trial state determinant can be obtained by writing the $M\times N$  matrix with orthogonal columns as a product of $(M-N)N$ Givens rotations \cite{Fishman_2015}
\begin{align}
    U \sim \prod_{i\in \text{occ},a\in\text{virt}} G_{ia}(\theta_{ia}) I_{M,N}\label{eq:givens}
\end{align}
where $I_{M,N}$ is a matrix consisting of the first $N$ columns of an identity matrix of size $M\times M$. 
When we say $A\sim B$ for two $M\times N$ matrices $A,\ B$, we mean that they represent the same SD up to an overall sign (or phase). More concretely, the two matrices are related to each other by orthogonal column operations or that $A A^T = B B^T$. The matrix $I_{M,N}$ corresponds to a determinant in which the first $N$ orbitals ($\vartheta_{i}$) are occupied. In this work, the underlying orbitals ($\vartheta_{i}$) are chosen as canonical Hartree-Fock orbitals arranged in increasing order with energy. Therefore, $I_{M,N}$ corresponds to the Hartree-Fock determinant, which is also used as the trial state $|\Psi_T\rangle$. 

The Givens rotations $G_{ia}(\theta)$ are defined as
\begin{equation}
    G_{ia}(\theta) = \begin{bmatrix}
        1&\cdots & 0 &\cdots &0& \cdots& 0\\
        \vdots & \ddots & \vdots & & \vdots &  & \vdots\\
        0 & \cdots & \cos(\theta) & \cdots & -\sin(\theta) & \cdots & 0\\
        \vdots &  & \vdots & \ddots & \vdots & & \vdots\\
        0 & \cdots & \sin(\theta) & \cdots & \cos(\theta) & \cdots & 0\\
        \vdots & & \vdots & & \vdots & \ddots & \vdots\\
        0 & \cdots & 0 & \cdots & 0 &\cdots & 1
    \end{bmatrix}
\end{equation}
where the trigonometric functions appear at the intersection of the $i$th and $a$th rows and columns, corresponding to rotations between occupied orbital $i$ and virtual orbital $a$.

When the Hartree-Fock determinant is transformed into another determinant using the Givens rotations, the order of rotations in (\ref{eq:givens}) must be fixed, since these matrices generally do not commute with each other. In our work, for every occupied index $i$ we loop over all virtual indices $a$. 

Below, we point out some facts about the Givens rotation based parametrization that are helpful in understanding and implementing our algorithm.
\begin{enumerate}
\item One can recover the space of singles, doubles, triples, $\cdots$ excitations commonly used in configuration interaction (CI) expansions, by fixing Givens rotation angles $\theta_{ia}$ to either $\pi/2$ or 0. For example, a determinant with a single excitation relative to HF, where an electron in orbital $i$ is excited to $a$, can be obtained by fixing all angles to 0 except for $\theta_{ia} = \pi/2$. Similarly, for higher-order excitation $\{i,j,\cdots\}\rightarrow\{a,b,\cdots\}$, one sets angles $\theta_{ia} , \theta_{jb}, \cdots$ to $\pi/2$ and the rest to 0. In total, there are $\binom{M}{N}$ distinct orthogonal determinants that span the Hilbert space. However, in AFQMC we work in the continuous space where each $\theta_{ia}$ can take any value in $(-\pi/2, \pi/2)$. This results in an overcomplete space of SDs, allowing the algorithm to explore the full manifold continuously rather than being restricted to discrete CI excitations. 

\item If any of the $\theta_{ia}$ in (\ref{eq:givens}) is equal to $\pm \pi/2$, then the resulting determinant is orthogonal to the reference (Hartree-Fock) determinant. By explicitly multiplying out all the Givens rotations acting on $I_{M,N}$, one finds that the resulting orthogonal matrix can be written in the block form: \begin{align}
    U = \begin{bmatrix}
    A\\
    B
\end{bmatrix}
\end{align} where $A$ and $B$ are block matrices of size $N\times N$ block and $(M-N)\times N$ respectively. The $A$ matrix is lower triangular with diagonal elements \[A_{i,i}=\prod_{a=N+1}^M\cos(\theta_{i,a}).\] 
Consequently, the overlap of the SD ($|\phi\rangle$) represented by $U$ with the trial state determinant represented by $I_{M,N}$ is given by
\begin{align}
    \langle \phi|\Psi_T\rangle =\prod_{i=1}^N \prod_{a=N+1}^M\cos(\theta_{i,a}).
\end{align}
Hence, if any of these angles are equal to $\pm \pi/2$, then the overlap vanishes: $\langle \phi|\Psi_T\rangle=0$. 

\item Not all SDs that are orthogonal to the reference determinants can be adequately represented using parametrization in (\ref{eq:givens}). For example, consider the case of a single electron in 3 orbitals. Then the determinant \[U=\begin{bmatrix}0,\\ 1/\sqrt{2}\\ 1/\sqrt{2}\end{bmatrix}\] cannot be obtained by any set of Givens rotations starting from $I_{3,1}$. However, any determinant that is not exactly orthogonal to the reference determinant but is arbitrarily close to an orthogonal determinant can be represented using the parametrization in $(\ref{eq:givens})$. For example the determinant \[U=\begin{bmatrix}\epsilon\\ \frac{1}{\sqrt{2}}\left(1-\frac{\epsilon^2}{2}\right)\\ \frac{1}{\sqrt{2}}\left(1-\frac{\epsilon^2}{2}\right)\end{bmatrix}\] can be parametrized with Givens rotations \[\theta_{12}=\frac{\pi}{4},~~ \theta_{13}=\frac{\pi}{2}-\sqrt{2}\epsilon\] correct up to $O(\epsilon)$.

\item For any matrix $U$ that has a non-zero overlap with $I_{M,N}$ (i.e. $\det[U^T \cdot I_{M,N}]\neq 0$), we can always obtain a set of Givens rotations $\theta_{ia} \in (-\pi/2, \pi/2)$ that satisfy (\ref{eq:givens}). To see this, write \[U=\begin{bmatrix}
    A\\
    B
\end{bmatrix}\] where $A,B$ are respectively the $N\times N$ and $(M-N)\times N$ blocks. In the first step, we perform an RQ-decomposition of $A$:
\begin{align}
    A = R_1Q_1
\end{align}
where $R_1$ is an upper triangular matrix and $Q_1$ has orthogonal columns.
The transformed matrix \[\tilde{U} = UQ_1 = \begin{bmatrix}
    R_1\\
    BQ_1
\end{bmatrix} \] represents the same SD as $U$ (up to an overall sign) because this operation is an orthogonal transformation of the columns of $U$. Because $U$ is assumed not orthogonal to $I_{M,N}$, all diagonal elements of $R_1$ are non-zero. Now we systematically remove virtual contributions in the lower block of $\tilde{U}$. Loop over occupied orbitals $i\in[1,N]$ in the reverse order and for each $i$, loop over all virtual orbitals $a\in[N+1,M]$ also in the reverse order. For each $(i,a)$ pair find
\begin{align}
    \theta_{ia} = \tan^{-1}\left(\frac{\tilde{U}_{ai}}{\tilde{U}_{ii}}\right),~~-\pi/2< \theta_{ia}<\pi/2
\end{align}
and update $\tilde{U} \leftarrow G_{ia}(-\theta_{ia})\tilde{U}$ which sets the $(i,a)$ element of $\tilde{U}$ to zero. If we perform this operation iteratively over all pairs of occupied and virtual orbitals, we will end up with $\tilde{U}$ that has zeros in the lower $(M-N)\times N$ block. This resulting matrix represents the same determinant as $I_{M,N}$ and thus we can write the equation
\begin{align}
    I_{M,N}&\sim \prod_{i=1}^N\prod_{a=N+1}^{M} G_{ia}(-\theta_{ia})U \\
    U&\sim \prod_{i=N}^1\prod_{a=M}^{N+1} G_{ia}(\theta_{ia})I_{M,N} 
\end{align}
where, to go from the first line to the second line, we use the fact that $G_{ia}(\theta) G_{ia}(-\theta) = I$.

\item Although we do not make use of it in this work, it is worth noting that one can fully parametrize the space of all Slater determinants (i.e., the Grassmann manifold) by introducing a collection of charts. Specifically, we require $\binom{M}{N}$ charts, each defined by choosing a different reference determinant in place of $I_{M,N}$ in (\ref{eq:givens}). This need arises because, as shown earlier, starting from a reference determinant, we cannot adequately parametrize determinants that are exactly orthogonal to it. Moreover, as pointed out before, there are $\binom{M}{N}$ mutually orthogonal determinants. 

The algorithm in the previous point can fail to find Givens rotations if $U$ is orthogonal to $I_{M,N}$. In this case we will start with a reference determinant $I_{M,\mathcal{I}}$ where $\mathcal{I} = \{i_1, i_2,\cdots,i_n\}$ not necessarily restricted to be the first $N$ columns. To find a suitable set of $\mathcal{I}$, we can perform a QR-decomposition with column pivoting on $U^T$ to obtain a permutation matrix with column indices arranged in decreasing order of importance. By selecting the first $N$ column index from the permutation matrix, we can find the indices $\mathcal{I}$ and thus the corresponding reference matrix $I_{M,\mathcal{I}}$. 

By using the same elimination algorithm as before, we can obtain rotations such that
\begin{align}
  U \sim \prod_{i\in\mathcal{I}}\prod_{a\notin\mathcal{I}} G_{ia}(\theta_{ia})I_{M,\mathcal{I}}  
\end{align}
\end{enumerate}
In the next section we will discretize (\ref{eq:fokker-planck}) where $\vec{\theta}$ will be a set of Givens rotation angles.

\subsection{Numerical solution of the Fokker-Planck equation}
To make the discussion concrete we will describe our algorithm for a simple system containing a single electron in three orbitals. This case corresponds to a two-dimensional Grassmann manifold parametrized by angles $\vec{\theta}=\{\theta_{12},\theta_{13}\}$. 

\subsubsection{Obtaining the parameters in Fokker-Plank equation}
As a first step we need to evaluate the three parameters $\mathcal{D}_{ij}, \mathcal{V}_i,\mathcal{S}$ that appear in the equation. These parameters are obtained from quantities 
\[g_i(\vec{\theta}), ~~ h_{ii}(\vec{\theta}), ~~ \vec{\gamma}_{i}(\vec{\theta}), ~~ \vec{\kappa}_{ij}(\vec{\theta}), ~~ w(\vec{\theta}), ~~ \vec{\chi}_{i}(\vec{\theta}),\] which we evaluate numerically. The first two are associated with the one-body operator and the last four with the two-body propagator. 

We first consider the two-body case. Acting with $\exp(x_k H_k)$ on the SD $|\phi(\vec{\theta})\rangle$ yields a non-normalized SD $|\gamma(x_k)\rangle$, represented by a matrix $A(x_k)$. Performing a QR-decomposition on $A(x_k)$ gives an upper-triangular matrix $R(x_k)$ and a matrix $Q(x_k)$ with orthogonal columns. The normalization weight is then \[w(x_k)=\prod_i R_{ii}(x_k),\] while the new set of Givens rotation parameters $\vec{\theta}(x_k)$ parameterizing $Q(x_k)$ are obtained by algorithm described in point 4 of section~\ref{sec:parametrization}.  Schmetially we can summarize this mapping as \[(w, \vec{\theta'}) = f(x_k, H_k, \vec{\theta}).\] 
We compute the first and second derivative of the outputs $w$ with respect to $x_i$ which gives us $g_i(\vec{\theta})$ and $h_{ii}(\vec{\theta})$ respectively, while the first and second derivative of $\vec{\theta'}$ with respect to $x_i$ gives us $\vec{\gamma}_{i}(\vec{\theta})$ and $\vec{\kappa}_{ij}(\vec{\theta})$ respectively. The derivatives are evaluated using algorithmic differentiation, as implemented in JAX. This process is repeated for each auxiliary field $x_i$. 

For the one-electron propagator the procedure is identical except that $x_k$ is replaced with $\tau$ and $H_k$ is replaced with $T_1$. 

With these first and second-order derivatives in hand, we can calculate the parameters $\mathcal{D}_{\alpha\beta}, \mathcal{V}_\alpha,\mathcal{S}$ in the Fokker-Plank equation as a function of $\vec{\theta}$. 
 
 \subsubsection{Discretizing and solving the Fokker-Plank equation}
 \begin{figure}[tbp]
\centering
\includegraphics[width=0.5\columnwidth]{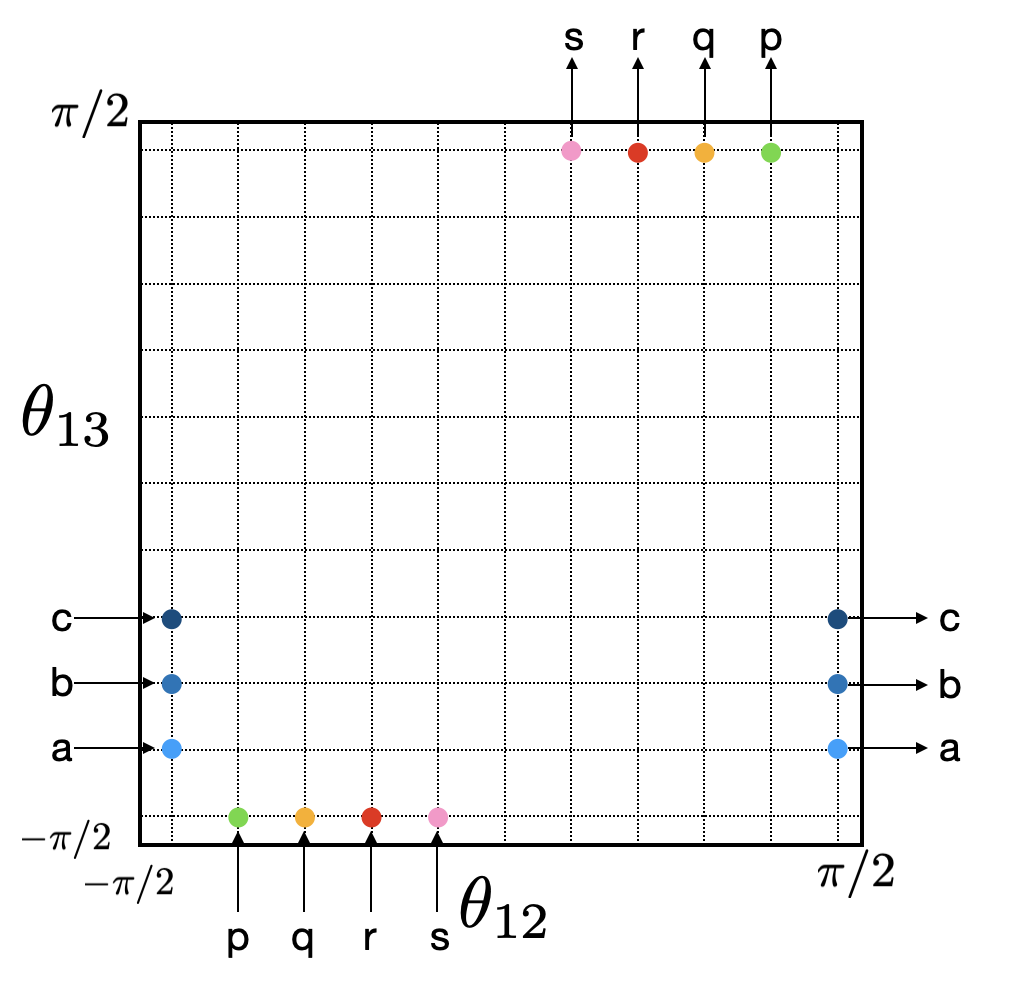}
\caption{
For a two-dimensional Fokker-Planck equation, we discretize the domain such that no grid points lie exactly on the boundary. The boundary conditions are represented schematically by the colored dots. Along the $\theta_{12}$ axis the wavefunction satisfies anti-periodic boundary condition $\phi(\pi/2+\delta, \theta_{13}) = -\phi(-\pi/2+\delta, \theta_{13})$ as indicated by the horizontal arrows on the left/right boundary points. In contrast, along the $\theta_{13}$ axis the wavefunction obeys a more unusual boundary condition: $\phi(\theta_{12}, \pi/2+\delta) = -\phi(-\theta_{12}, -\pi/2+\delta)$ which is illustrated by the vertical arrows.
}\label{fig:BC}
\end{figure}

 Once the parameters in the PDE have been obtained, we discretize the space of Givens rotations using a uniformly spaced grid of $N_{g}$ points along each coordinate. The grid values is given by
 \begin{align}
     \theta_\alpha(p) = \frac{(1-N)\pi}{2N}+\frac{p\pi}{N}
 \end{align}
 for $p=0,1,\cdots,N-1$. Note that this places $N_g$ grid points between $-\pi/2$ to $\pi/2$ but avoids placing any points exactly on the boundaries where parametrization becomes redundant. By systematically increasing $N_g$, we can get arbitrarily close to the boundary. The choice of grid effectively circumvents the issue that the Givens rotation parametrization cannot represent determinants that are exactly orthogonal to the reference determinant. While this is an ad-hoc fix, we will demonstrate through numerical experiments that increasing $N_g$ yields results that converge to those obtained from the AFQMC algorithm formulated in terms of Langevin dynamics. 

With the grids, we approximate the first and second partial derivatives using central differences:
 \begin{align}
     \frac{\partial^{2}\psi}{\partial \theta_{\alpha}\partial \theta_{\beta}}(\cdots,p,\cdots,q,\cdots) \approx& \frac{\left(\psi(\cdots,p-1,\cdots,q-1,\cdots)-\psi(\cdots,p-1,\cdots,q+1,\cdots)\right.}{}\nonumber\\
     &\frac{\left.-\psi(\cdots,p+1,\cdots,q-1,\cdots)+\psi(\cdots,p+1,\cdots,q+1,\cdots)\right)}{\Delta p\Delta q}\nonumber\\
         \frac{\partial\psi}{\partial \theta_{\alpha}}(\cdots,p,\cdots,q,\cdots) \approx& \frac{\psi(\cdots,p+1,\cdots,q,\cdots)-\psi(\cdots,p-1,\cdots,q,\cdots)}{2\Delta p}
 \end{align}
 where $\psi(p,q,\cdots)$ is a short hand for the value of the function at grid points $\psi(\theta_{1}(p), \theta_{2}(q),\cdots)$. 

The finite-difference scheme can be applied to all points within the interior of the domain. At the boundary, however, one must impose suitable boundary conditions.
\begin{enumerate}
    \item \textbf{Free Projection:} For free projection we use the finite difference scheme at the boundary but must decide how to handle points outside the domain. Such points must be mapped back into the domain. This mapping is nontrivial due to the unusual topological structure of the Grassmann manifolds. The boundary condition is neither periodic nor anti-periodic as illustrated in Figure~\ref{fig:BC}. In higher dimensions no closed-form analytic expression of the form given in caption of Figure~\ref{fig:BC} is known to us, because of the non-commuting nature of the Givens rotations. Instead we determine it numerically. Suppose the $\alpha^{th}$ coordinate is close to the boundary: $\theta_\alpha = \pi/2-\delta$. To find the corresponding interior point for $\theta_{\alpha} = \pi/2+\delta$ (which takes us outside the domain) we construct a set of Givens rotations with $\theta_\alpha$ lying outside the domain and form the corresponding matrix $U$. We then apply the algorithm in Sec.~\ref{sec:parametrization}, point 4, to obtain a new set of parameters representing the same matrix $U$ up to an overall sign.
\item \textbf{Constraint path:} For cp-AFQMC the boundary condition is simpler: we enforce that all Slater determinants orthogonal to the trial state have zero weight. This amounts to putting Dirichlet boundary condition, \[\psi(\vec{\theta})=0\quad \text{if any } \theta_\alpha = \pm \pi/2.\] 
Since our discretization avoids placing points exactly at the boundary, we enforce $\psi=0$ for any grid point next to the boundary i.e. $\theta_\alpha(p=0) = \theta_\alpha(p=N-1)=0$. This is not exactly equivalent to the constraint in cp-AFQMC but gives results that converge to the cp-AFQMC limit as $N$ is increased.
\end{enumerate}

 Once the Fokker-Planck equation has been discretized with an appropriate boundary condition the time-dependent Fokker-Planck equation takes the form
 \begin{align}
     \frac{d\psi}{d\tau} = F\psi 
 \end{align}
 where $\psi$ is a vector of size $N_{g}^{N_\theta}$, where $N_{\theta}$ is the number of paramaeters and $N_g$ is the number of grid points along each coordinate, $F$ is the matrix obtained by discretizing the Fokker-Planck equation. The matrix is sparse and only has $2N_\theta+1$ number of non-zero elements per row. 
 
 Because $\mathcal{D},\mathcal{V}$ are position dependent, $F$ is not symmetric. In the large-$\tau$ limit the solution corresponds to the eigenvector of $F$ with the eigenvalue having the smallest real part. This state can be obtained by propagating in imaginary time with a sufficiently small time step or directly by diagonalizing the matrix $F$ both of which give the same result. The eigenvalues themselves do not have any physical meaning, however the eigenvectors correspond to the physical state defined in (\ref{eq:state}).

\section{Results}\label{sec:results}
The simplest system in which we have observed a bias in the wavefunction is a 3-site Hubbard model with a single electron. We will begin by analyzing this system and then move on to 2-electron systems. One of the aims is to demonstrate that by exactly solving with Fokker-Planck equation with sufficient number of grid points we are able to reproduce the free projection and constrained path results. These simulations will help us understand the effect of the constraint on the solution. We note that all AFQMC results here are converged with respect to the time step error and the number of walkers. The bias in the energies is smaller than the stochastic noise of the results. Also we note that we have used a discrete Hubbard-Stratonovich decomposition instead of the continuous one shown in (\ref{eq:HScont}) for AFQMC calculations, however, both are known to give the same results. 

Before presenting the results, we note that two different wavefunctions are used in these calculations: one to impose the constraint (guiding wavefunction) and the other to evaluate the projected energy (trial wavefunction). In all examples that follow, the trial state is obtained by applying Givens rotations with small random angles to the guiding wavefunction. Usually these two wavefunctions are chosen to be identical, however there are instances in the following examples where the guiding wavefunction is exact and if the same wavefunction is used for evaluating the projected energy then by zero-variance principle we are guaranteed to have no bias in the energy, even if the AFQMC wavefunction is not exact. To diagnose the bias in the wavefunction, we deliberately use an inexact trial state, so that any error in the projected energy reflects the bias in the AFQMC wavefunction. 

\subsection{One-electron in three orbitals}
\begin{figure}
    \centering
    \includegraphics[width=0.9\linewidth]{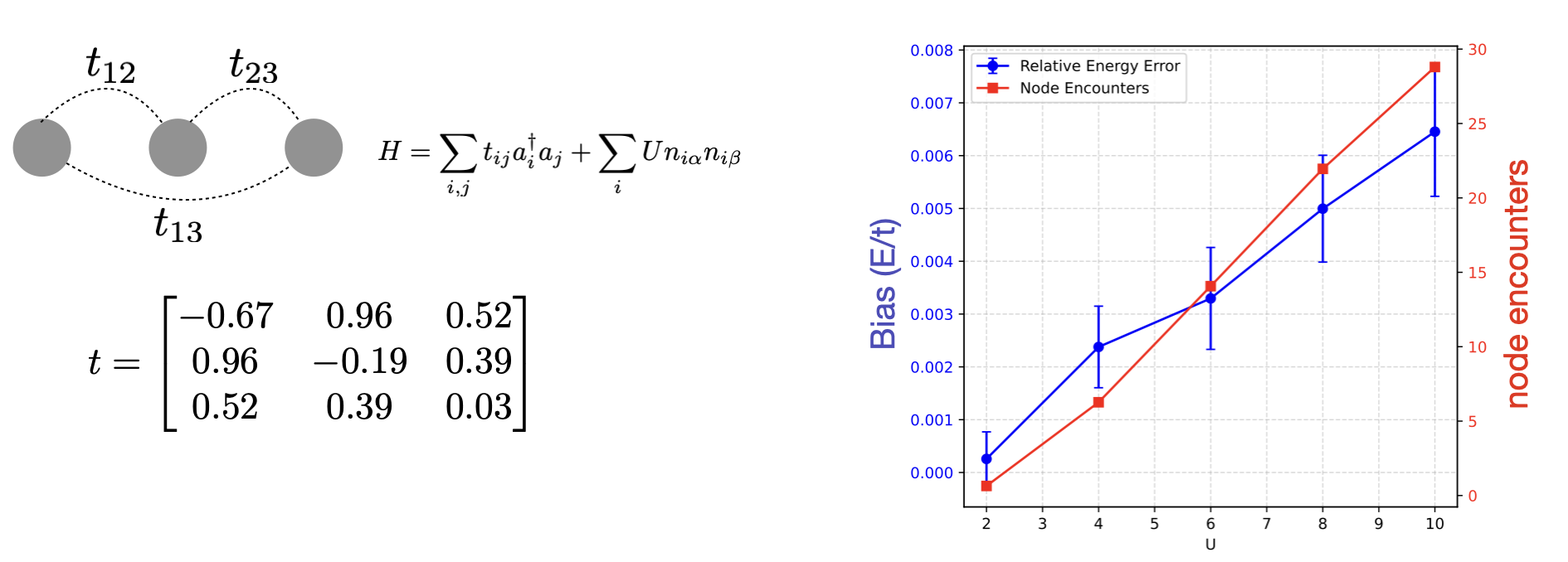}
    \caption{The figure shows the Hubbard model on the left with a general one-electron matrix. On the right we show the result of constraint path AFQMC calculations as a function of increasing $U$ with the ground state of the Hamiltonian as the guiding wavefunction. We also show the number of node encounters per imaginary time step which is strongly correlated with the bias in energy.}
    \label{fig:1e3o_system}
\end{figure}
The system is schematically shown in Figure~\ref{fig:1e3o_system}. Since the system constraints only a single electron, the mean field result is exact. The two-body operation, that describes interaction between $\alpha$ and $\beta$ electrons on the same site, is inactive in this system. However, after Hubbard-Stratonovich transformation is performed the instantaneous action of the two-body term becomes non-zero in both cp-AFQMC and Fokker-Planck formulation. Consequently, the diffusion and drift contribution from the two-electron operator are non-zero in the latter. 

When the Hartree-Fock state (which is the exact ground state of the Hamiltonian) is used as the guiding wavefunction in cp-AFQMC, the projected energy exhibits a bias that increases with increasing $U$. Interestingly, the number of node encounters also increases with $U$ and is strongly correlated with the bias in the energy. It is worth mentioning that the number of node encounters is non-zero even in the small time limit because we are not using importance sampling. The use of importance sampling does not alter the ph/cp-AFQMC energy. 

\begin{figure}[ht]
    \centering
    \includegraphics[width=0.8\linewidth]{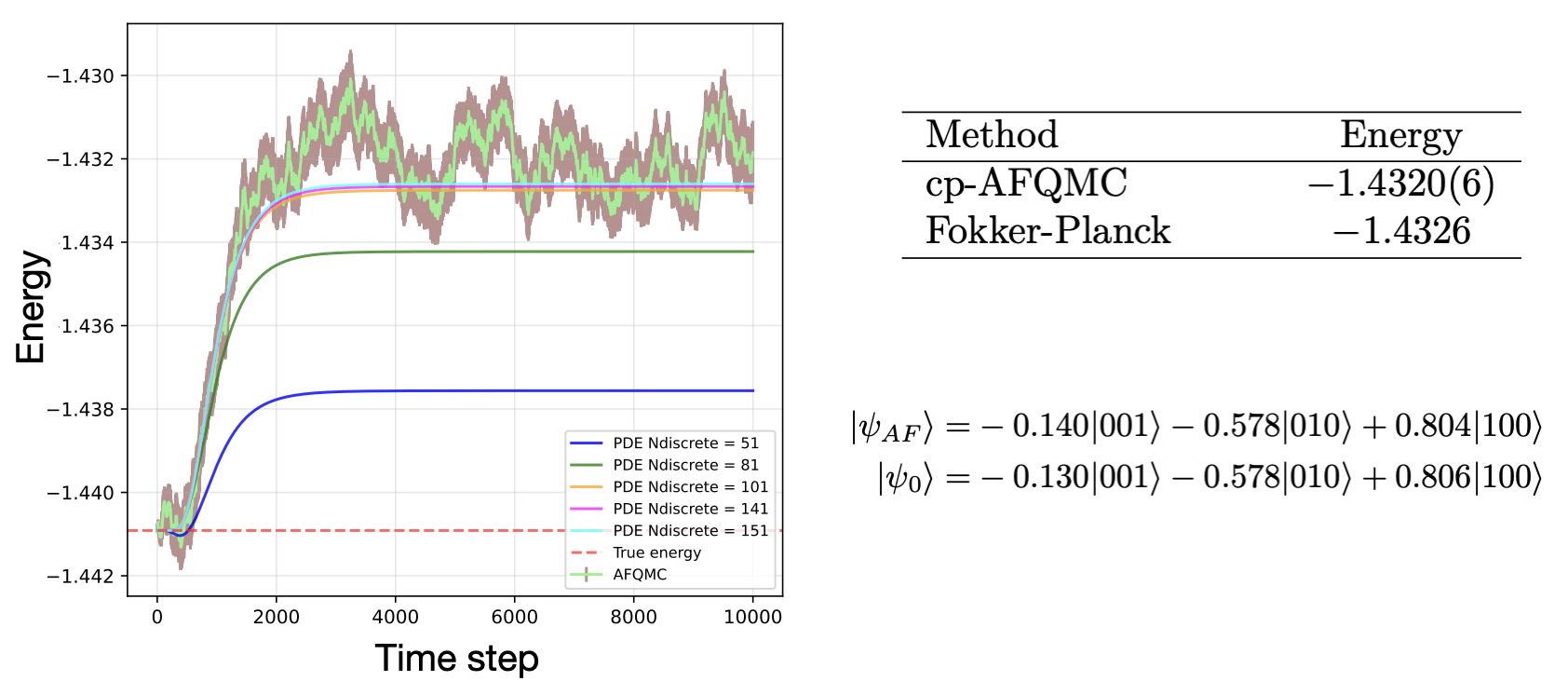}
    \caption{The graph shows the convergence of Fokker-Planck results with increasing discretization using solids curves. The dotted orange lines is the exact ground state energy and the green lines with brown error bars correspond to the results from cp-AFQMC simulation. The table on the right shows that the converged energy from Fokker-Planck equation agrees with the cp-AFQMC simulations up to an error bar. The exact state ground state ($|\psi_0\rangle$) and the cp-AFQMC states ($|\psi_{AF}\rangle$) are shown below the table in site basis.}
    \label{fig:1e3o}
\end{figure}
We first solve the Fokker-Planck equation derived in the previous section using the Dirichlet boundary condition that are designed to mimic the cp-AFQMC simulation with $U=10$, a case for which cp-AFQMC exhibits a bias. The solution $\psi$ is a function of two parameters $\theta_{12},\theta_{13}$ and we converge the results with respect to the number of grid points $N_g$
along each coordinate. Figure~\ref{fig:1e3o} shows that with increasing $N_g$ the Fokker-Planck results agree with the cp-AFQMC results up to stochastic noise. In addition to the converged results we note that the transient results of the two simulations also agree with each other. In both cases, the calculation is initialized with Hartree-Fock determinant at $\tau=0$, which corresponds to  $\psi(\vec{\theta}, \tau=0) = \delta_{\vec{\theta}}$ in Fokker-Planck equation (in practice, the delta function is replaced by a discretized approximation). Consequently, the initial energies are exact and remain constant for a short imaginary time while the function $\psi(\vec{\theta})$ evolves, remaining confined in the interior of the domain ($\theta_\alpha \in(-\pi/2, \pi/2)$). Note, that the Fokker-Planck equation with Dirichlet boundary conditions behaves exactly like free projection as long as the $\psi$ remains zero near the boundaries. As the state propagates forward, $\psi$ spreads out and approaches the boundaries, where the constraint begins to act. This is when the bias in energy start to emerge. 

\begin{figure}[hbt]
    \centering
    \includegraphics[width=0.8\linewidth]{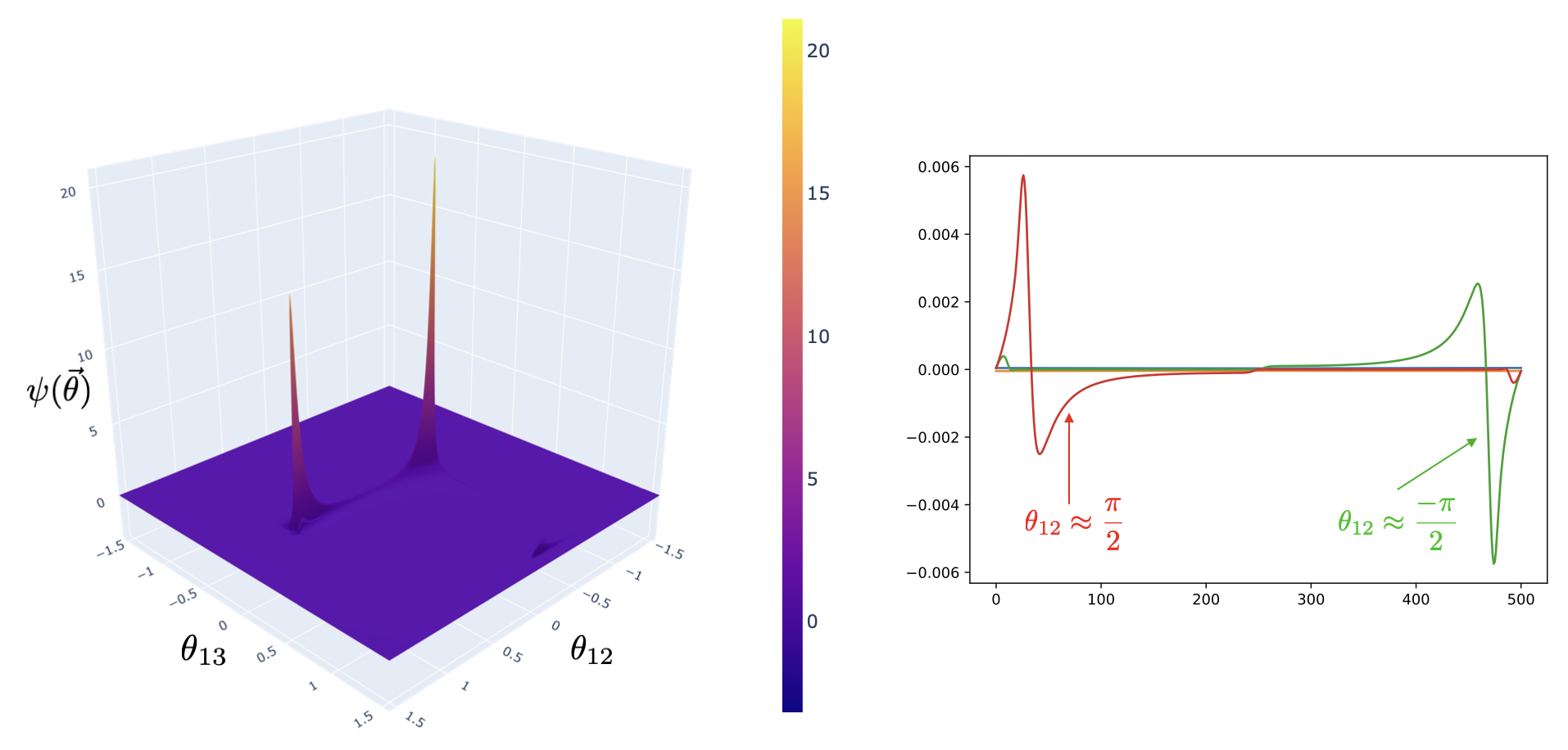}
    \caption{The left panel shows the surface plot of $\psi(\vec{\theta})$ and the right plot shows its values on the two boundaries. Note that the function remains non-zero on the two boundaries. The value of the function at $\theta\approx\frac{\pi}{2}$ is negative and a mirror image of its value at $\theta\approx-\frac{\pi}{2}$ reflecting the unusual boundary condition of the Grassmann manifold.}
    \label{fig:placeholder}\label{fig:1e3o_2}
\end{figure}

\begin{figure}[hbt]
    \centering
  \includegraphics[width=0.49\linewidth]{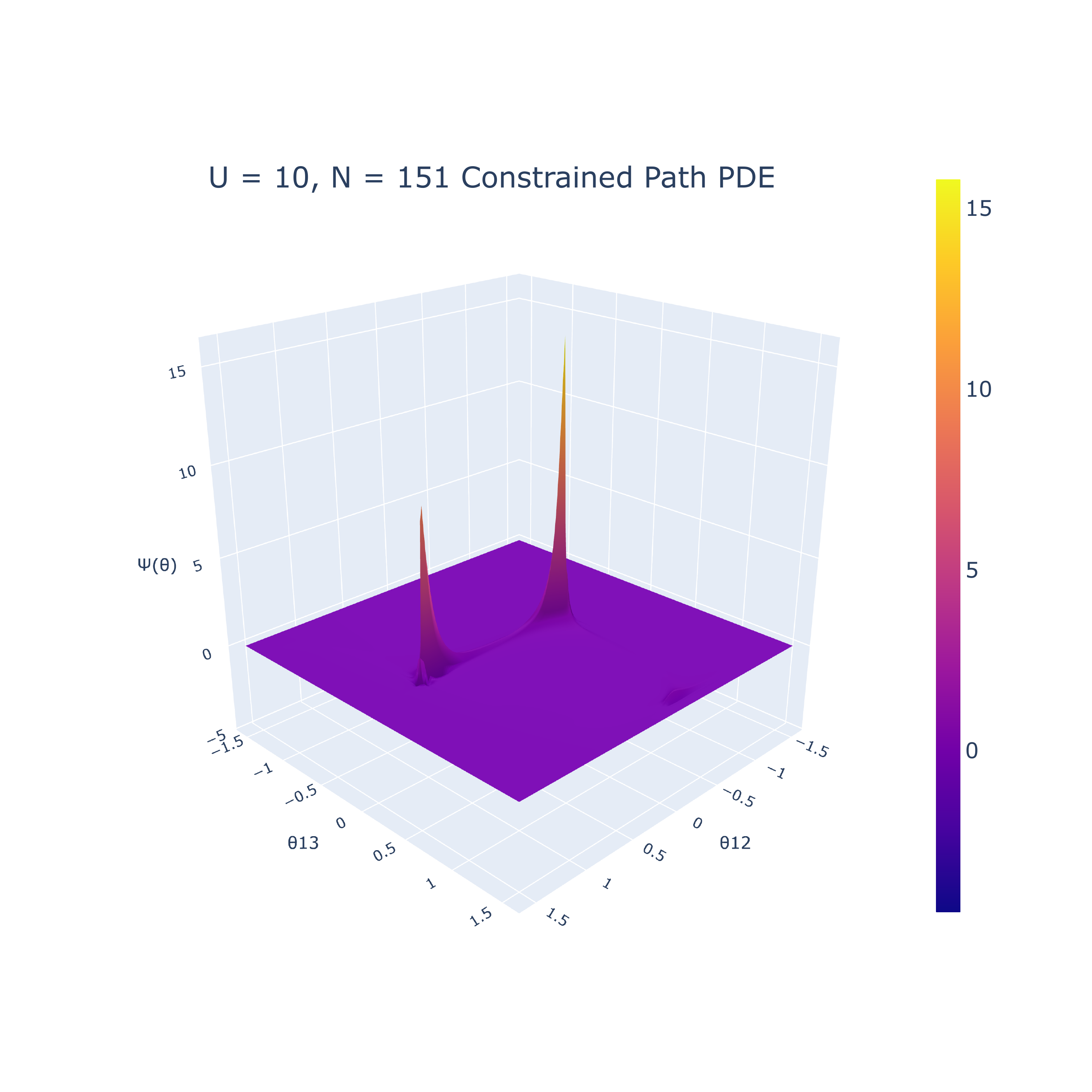}
  \begin{picture}(0,0)
    \put(-220,205){\footnotesize (a)}
  \end{picture}\
    \includegraphics[width=0.49\linewidth]{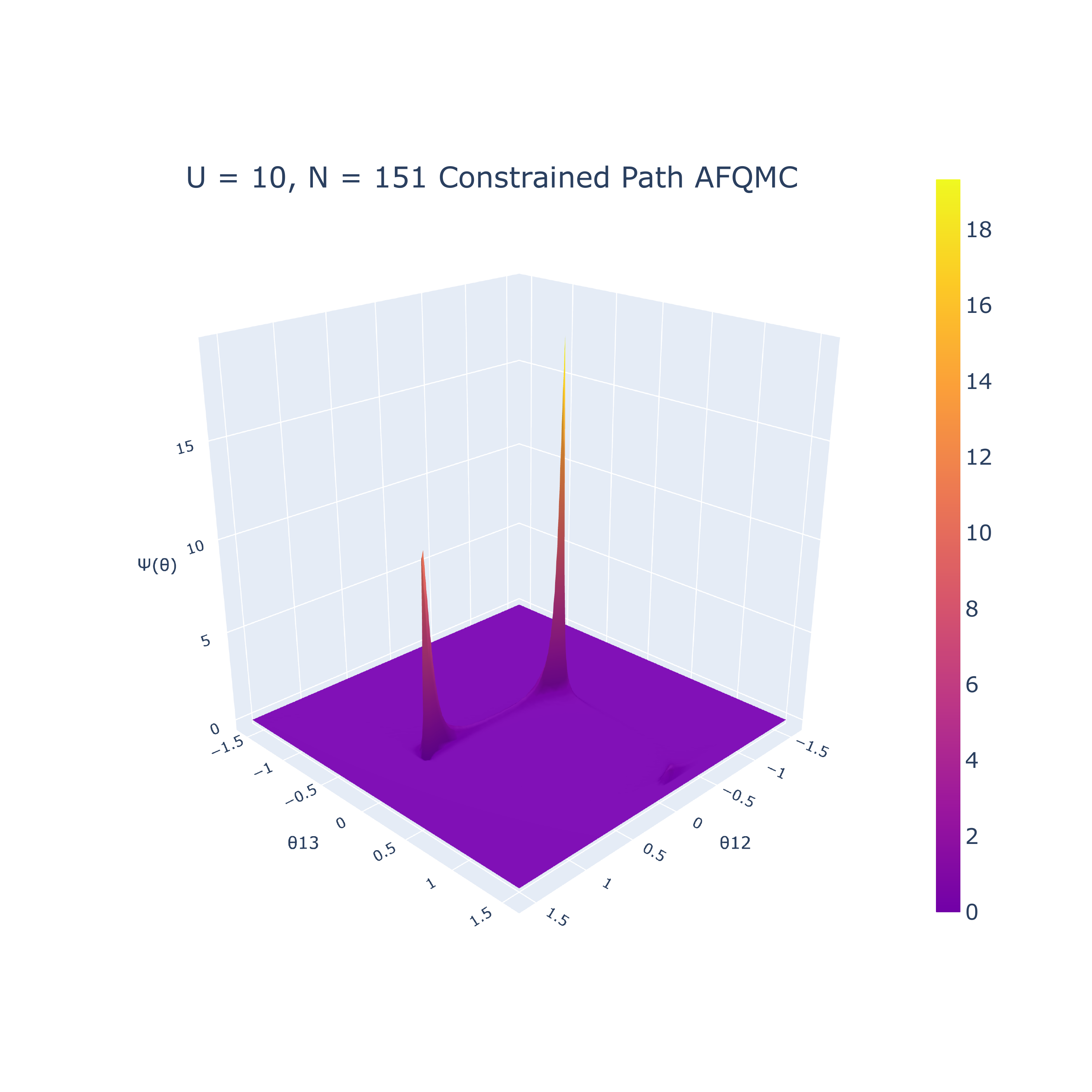}
  \begin{picture}(0,0)
    \put(-220,205){\footnotesize (b)}
  \end{picture}\
    \caption{Surface plots of wavefunction $\psi_{\text{PDE}}(\vec{\theta})$ derived by Fokker-Planck Equations with Dirichlet Boundary conditions (a) and the distribution of walkers $\psi_{\text{AFQMC}}(\vec{\theta})$ sampled by cp-AFQMC (b). Qualitatively, we see a remarkable degree of similarity between the two including not only the bimodal distribution, but even the tiny peak near the boundary.}
    \label{fig:placeholder}
\end{figure}

Next we solve the Fokker-Planck equation without any constraints with $N_g=501$. The converged solution $\psi(\vec{\theta})$ is show in Figure~\ref{fig:1e3o_2}. Interestingly, the value of $\psi$ is non-zero on the boundaries which correspond to determinants that are exactly orthogonal to the ground state. This indicates that in cp-AFQMC, the walkers should indeed sample determinants from the nodal surface (i.e. states orthogonal to the ground state). In cp-AFQMC we set the weights of these walkers to zero, which introduces a bias. It is worth noting that for a smaller $U$ the diffusive term is weaker and $\psi(\vec{\theta})$ has reduced tendency to approach the boundary. This leads to fewer node crossing and a smaller accompanying bias. This observation helps to explains that for problems with small effective $U$ even when Hartree-Fock is inexact cp-AFQMC can provide accurate results if $\psi$ does not approach the boundaries. However, our results also imply that one might not get uniform convergence towards the exact result as the quality of the trial state is improved. In fact we have observed cases in which the cp-AFQMC wavefunction gets poorer when the guiding wavefunction is variationally improved.  

As an additional check to demonstrate that the Fokker-Planck equation is reproducing the BRW distribution of AFQMC, we show in Figure~\ref{fig:placeholder}, the distribution $\psi_{\text{AFQMC}}(\vec{\theta})$ sampled by the walkers in cp-AFQMC and $\psi_{\text{PDE}}(\vec{\theta})$ the solution to the Fokker-Planck equation with $N_{g} = 151$ and Dirichlet boundary conditions. We can quantify the difference between the two distributions by calculating $\int(\psi_{\text{PDE}}(\vec{\theta})-\psi_{\text{AFQMC}}(\vec{\theta}))^{2}d\vec{\theta} \approx 0.14$. It should be noted that a notrivial amount of this difference comes from the fact that the PDE solution has negative values at certain $\vec{\theta}$, which is a byproduct of finite-difference and thus a numerical artifact. The strong qualitative agreement between the two distributions lends even stronger support to the fact that the Fokker-Planck equation is reproducing what is happening in AFQMC.

The results clarify the flaw in the argument outlined in Figure~\ref{fig:nodal}. The argument would hold if the space in which the walk was being performed was orthogonal (as is the case in diffusion Monte Carlo). In the case of cp-AFMQC the walk is performed in the space of overcomplete manifold of Slater determinants, where, the overlap with the trial state varies smoothly and vanishes at the node. The contribution from determinants exactly orthogonal to the ground state can be canceled by other determinants in this manifold, since they are not mutually orthogonal.

\subsection{Two-electron system}
We now perform calculations on two-electron systems by solving the Fokker-Planck equations. Going to larger system sizes with a deterministic algorithm is challenging because of the rapid increase in the dimension of the problem. 

\begin{figure}[tb]
    \centering
    \includegraphics[width=0.8\linewidth]{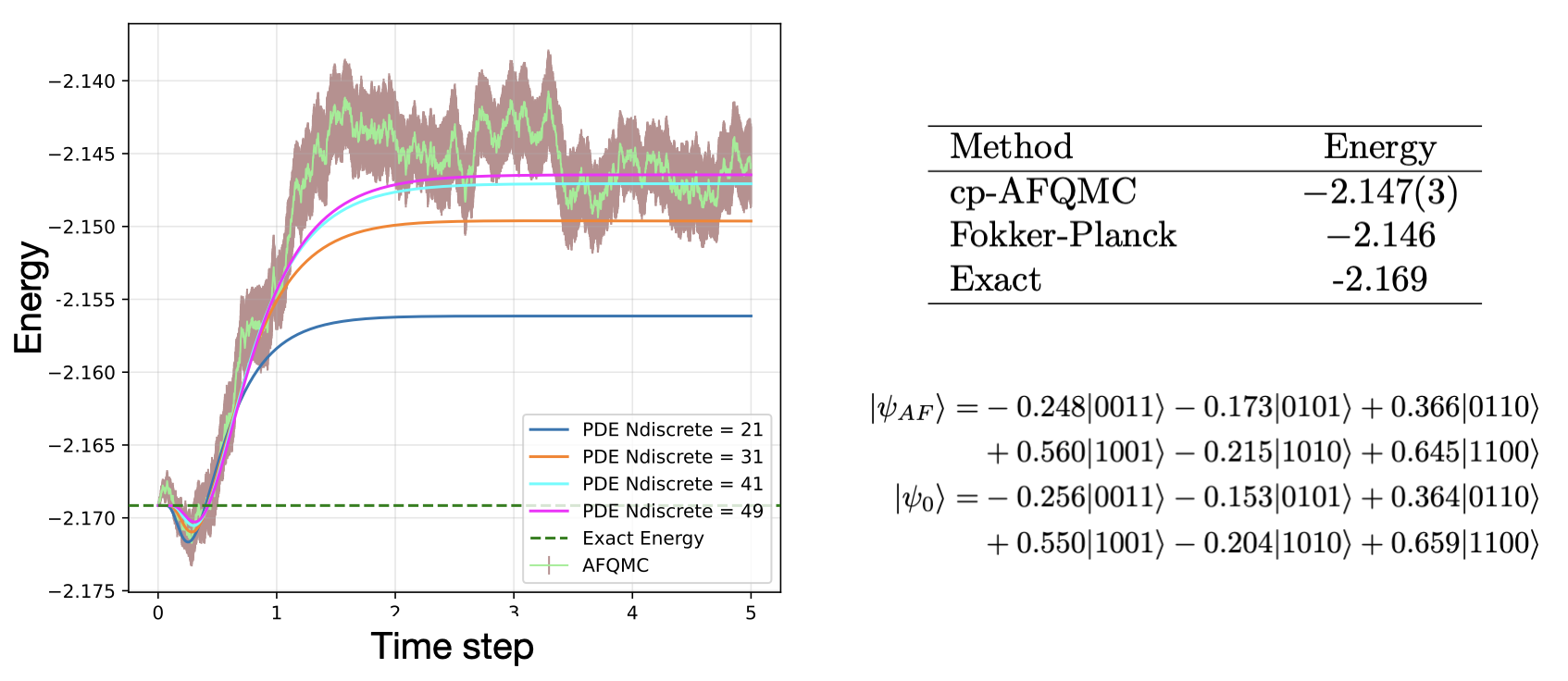}
    \caption{With sufficiently fine discretization, the solution of the Fokker–Planck equation reproduces the cp-AFQMC results for system containing four orbitals with two $\alpha$ electrons. The table shows that the converged energies from Fokker-Planck equations agree with cp-AFQMC up to stochastic noise and both these results are biased relative to exact results despite the trial state being exact. The exact state ground state ($|\psi_0\rangle$) and the cp-AFQMC states ($|\psi_{AF}\rangle$) are shown below the table in site basis.}
    \label{fig:2a4o}
\end{figure}
As a first example, we consider a system containing 4 orbitals and 2 $\alpha$ electrons. This system also has no effective two-body interaction, so the Hartree-Fock state is the exact ground state. When cp-AFQMC is performed for this system with the Hartree-Fock state as a guiding wavefunction, we observe a trend that is similar to single electron case of the previous section as shown in Figure~\ref{fig:2a4o}. There is a short initial time range where the energy remains exact and then it evolves to a biased result. The agreement between cp-AFQMC and Fokker-Planck equation is again better than stochastic noise. 

\begin{figure}[htb]
    \centering
    \includegraphics[width=0.8\linewidth]{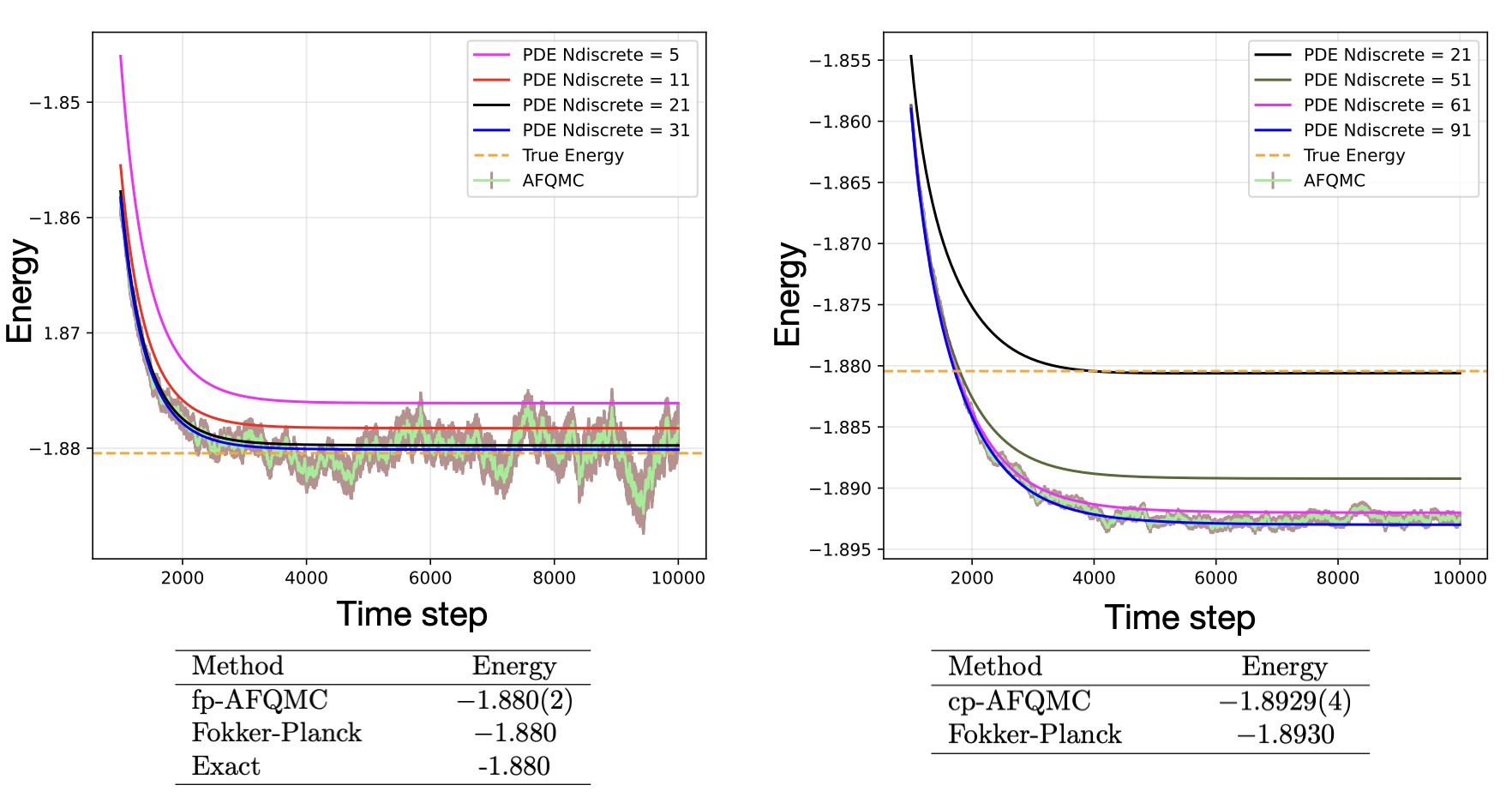}
    \caption{With sufficiently fine discretization, the solution of the Fokker–Planck equation reproduces the fp-AFQMC results (left panel) as well as the cp-AFQMC results (right panel) for system containing three orbitals with one $\alpha$, one $\beta$ electron and a Hubbard-$U=3$. The corresponding exact ground-state energies are summarized in Table at the bottom of the graphs, and demonstrate quantitative agreement.}
    \label{fig:11e4o}
\end{figure}
Next we example a 3-orbital system with one $\alpha$ and one $\beta$ electron. For this system the Hartree-Fock state is no longer the exact ground state. Our goal is to demonstrate that the results from the Fokker-Planck equation agrees with the cp-AFMQC and free projection AFQMC results even when non-zero number of electrons of both spins are present in the system and the guiding wavefunction is not exact. The results from fp-AFQMC and cp-AFQMC results are shown in Figure~\ref{fig:11e4o} along with tabulated energies for the two types of calculations.

\section{Conclusions}\label{sec:conclusions}
In this article we have derived and solved the Fokker-Planck equation that governs the populations of walkers sampled by cp-AFQMC and fp-AFQMC. When solved exactly, the Fokker-Planck equation yields the underlying AFQMC wavefunction as given in (\ref{eq:state}). Our analysis has been limited to the Hubbard model (it can be readily generalized to ab-initio Hamiltonians with diagonal two-electron integrals) where both the auxiliary fields and the wavefunction amplitudes are real numbers. We outline a few conclusions that can be drawn from this work.
\begin{itemize}
\item The converged state $\psi(\vec{\theta})$ is not unique for a given many-body wavefunction. For example, in the case of single electron in three orbitals, we find that different values of the Hubbard interaction $U$ lead to distinct solutions of the Fokker–Planck equation, even though the ground state itself is unchanged. This highlights how the exact form of the Hubbard-Stratonovich transformation can influence the results under constraint path. 

\item The cp-AFQMC amounts to imposing the condition that the solution of Fokker-Planck equation vanishes for all walkers that are exactly orthogonal to the trial state. In the current work, our trial state is always a single Slater determinant. In this case, the boundary of the domain parametrized by Givens rotations is exactly orthogonal to the reference state and thus the cp-AFQMC amounts to imposing the Dirichlet boundary condition. In AFQMC the nodal structure is imposed in determinant space, not in real space, so ``nodes of the guiding wavefunction" here refers to where overlaps vanish in the manifold coordinates, not in coordinate space. Contrary to what one might expect, the solution of Fokker-Planck equation without the Dirichlet boundary condition does not vanish on the nodes of the guiding wavefunction, even when the guiding wavefunction is exact. 

\item The constraint itself is not uniquely defined for a state $|\psi\rangle$ because its representation $\psi(\vec{\theta})$ on the Grassmann manifold is not unique. The constraint is only active on the boundaries of the manifold. This implies that if the fixed point of the Fokker-Planck equation has the property that it goes to zero sufficiently rapidly as one moves towards the boundary, then the constraint path will not introduce any bias. 
    
\item The constraint path maintains a positive $\psi(\vec{\theta})$ is the Grassmann manifold. An interesting question that one can ask is as follows. Can one always represent an arbitrary state with a positive $\psi(\vec{\theta})$ in the manifold? Although we cannot provide a rigorous proof for or against this proposition, we point out that it is quite plausible that the answer to the question posed is yes. To see this, note that the points near the boundary have an image on another boundary which represents the same determinant with an opposite sign. Also the determinants in the Hilbert space are represented by points on the boundary. Thus any CI expansion can be represented as a linear combination of points near the boundary with a positive weight. In other words $\psi(\vec{\theta})$ can be chosen as a linear combination of delta functions in our manifold to represent any CI expansion. In our manifold the points exactly on the boundary are not included but we can approach arbitrarily close to such points. However a complimentary question is, can constraint path ever have such a state as its fixed point. It is quite likely that this is not the case because the constraint requires that the wavefunction needs to smoothly approach 0 at the boundaries and the inclusion of constraint might never allow one to obtain such a $\psi(\vec{\theta})$. However, the proposed state is likely only one of many states and it is not clear if some of these states can be the fixed point of constraint path. 

\item In the current article we have not employed importance sampling. For cp-AFQMC, importance sampling leads to a modified representation of the wavefunction, 
\begin{align}
    |\psi(\tau)\rangle = \int \psi(\vec{\theta}, \tau) \langle\phi(\vec{\theta})|\psi_T\rangle |\phi(\vec{\theta})\rangle d\vec{\theta} \label{eq:imp}
\end{align}
rather than (\ref{eq:state}). Because of the over-completeness of the Slater determinant manifold, 
\[|\psi_T\rangle \neq \int  \langle\phi(\vec{\theta})|\psi_T\rangle |\phi(\vec{\theta})\rangle d\vec{\theta}\] 
and thus there appears to be no guarantee that $\psi(\vec{\theta},\tau)$ in (\ref{eq:imp}) exhibits reduced variance compared to (\ref{eq:state}). In fact, (\ref{eq:imp}) suggests that using importance sampling will lead to a biased AFQMC wavefunction even when free projection is used with an exact trial state. This is because this particular flavor of importance sampling always imposes the constraint, which therefore leads to a bias.
\end{itemize} 

The current work opens up several avenues for improving AFQMC. From the perspective of Fokker-Planck equation the Dirichlet boundary imposed in cp-AFQMC is only one possible choice, other approximations such as periodic boundary conditions may be explored. Another promising direction is to approximate $\psi(\vec{\theta})$ directly using machine learning methods such as neural networks or Gaussian processes regression. Such approaches require appropriate metric on the Grassmann manifold. The simple Euclidean metric will be inadequate because points that appear distant under Euclidean metric can correspond to nearly identical determinants, due to the unusual boundary conditions. If one could adequately parametrize the cp-AFQMC wavefunction $\psi_{CP}(\vec{\theta})$ which is already close to the ground state, then one could use perturbation theory to correct its value. The constraint can act like the perturbation and its action on the wavefunction can be corrected up to the first order. Another possibility is to search for Hubbard-Stratonovich transformation that preferentially give rise to diffusive and drift terms that tends to move the state away from the boundary towards the center. 

\section{Acknowledgement}
We would like to thank Garnet Chan, Michael Lindsey, Shiwei Zhang, and Miguel Morales for helpful discussions. This material is based upon work supported by the U.S. Department of Energy, Office of Science, Accelerated Research in Quantum Computing Centers, Quantum Utility through Advanced Computational Quantum Algorithms, grant no. DE-SC0025572.
\bibliography{references.bib}
\appendix
\section{Derivation of Fokker-Planck equation due to two-body operator}\label{eq:app1}
In (\ref{eq:twoBody1}) in step $(ii)$ let us look at the term that is linear in $\delta\tau$ in the integrand which is given below
\begin{align}
&-(\partial_\alpha \psi) (\gamma_{\alpha i}g_i-\gamma_{\alpha i}(\partial_\beta \gamma_{\beta i}) + \eta_{\alpha ii})+\frac{1}{2}(\partial_\alpha\partial_\beta\psi)\gamma_{\alpha i}\gamma_{\beta i}\nonumber\\ 
&+ \psi \left(h_{ii} - g_i (\partial_\alpha \gamma_{\alpha i}) - (\partial_\alpha g_i)\gamma_{\alpha i}-\partial_\alpha\eta_{\alpha ii}+\frac{1}{2}(\partial_\alpha \gamma_{\alpha i})(\partial_\beta \gamma_{\beta i})-\frac{1}{2}(\partial_\alpha \gamma_{\beta i}\partial_\beta \gamma_{\alpha i})\right)\nonumber   
\end{align}
Now we substitute \[\eta_{\alpha ii} = \kappa_{\alpha ii} - (\partial_{\beta}\gamma_{\alpha i})\gamma_{\beta i}\]
into the equation to get
\begin{align}
&-(\partial_\alpha \psi) (\gamma_{\alpha i}g_i-\gamma_{\alpha i}(\partial_\beta \gamma_{\beta i}) + \kappa_{\alpha ii} - (\partial_{\beta}\gamma_{\alpha i})\gamma_{\beta i})+\frac{1}{2}(\partial_\alpha\partial_\beta\psi)\gamma_{\alpha i}\gamma_{\beta i}\nonumber\\ 
&+ \psi \left(h_{ii} - g_i (\partial_\alpha \gamma_{\alpha i}) - (\partial_\alpha g_i)\gamma_{\alpha i}-\partial_\alpha\kappa_{\alpha ii} + \gamma_{\beta i} \partial_\alpha \partial_\beta \gamma_{\alpha i} + (\partial_{\beta}\gamma_{\alpha i})(\partial_\alpha \gamma_{\beta i})+\frac{1}{2}(\partial_\alpha \gamma_{\alpha i})(\partial_\beta \gamma_{\beta i})-\frac{1}{2}(\partial_\alpha \gamma_{\beta i}\partial_\beta \gamma_{\alpha i})\right)\nonumber    \\
=& -\partial_\alpha(\psi (\gamma_{\alpha i}g_i + \kappa_{\alpha ii})) + (\partial_\alpha \psi) (\gamma_{\alpha i}(\partial_\beta \gamma_{\beta i}) + (\partial_{\beta}\gamma_{\alpha i})\gamma_{\beta i})+\frac{1}{2}(\partial_\alpha\partial_\beta\psi)\gamma_{\alpha i}\gamma_{\beta i}\nonumber\\ 
&+ \psi \left(h_{ii} + \gamma_{\beta i}\partial_\alpha \partial_\beta \gamma_{\alpha i} +\frac{1}{2}(\partial_\alpha \gamma_{\alpha i})(\partial_\beta \gamma_{\beta i})+\frac{1}{2}(\partial_\alpha \gamma_{\beta i}\partial_\beta \gamma_{\alpha i})\right)\nonumber    \\
=&-\partial_\alpha(\psi (\gamma_{\alpha i}g_i + \kappa_{\alpha ii})) +\frac{1}{2}\partial_\alpha\partial_\beta (\psi \gamma_{\alpha i}\gamma_{\beta i}) + \psi h_{ii}
\end{align}

\section{Hopping matrix}\label{eq:app2}
Hopping Matrix for 3 orbitals, 1-alpha electron, 1-beta electron is the same as the 3 orbital, 1-alpha electron, 0-beta electron case:

\begin{equation}
\begin{pmatrix}
-0.6680644 & 0.9605467 & 0.520678\\

0.9605467 & -0.1900202 & 0.3938467\\

0.520678 & 0.3938467 & 0.0340779
\end{pmatrix}
\end{equation}

Hopping matix for 4 orbitals, 2-alpha electron, 0-beta electron is:
\begin{equation}
\begin{pmatrix}
-0.6680644 & 0.9605467 & 0.520678 & -0.1044906 \\
0.9605467 & -0.1900202 & 0.3938467 & -1.0511902 \\ 
0.520678 & 0.3938467 & 0.0340779& -0.4899371\\
-0.1044906 & -1.0511902 & -0.4899371 & 0.0882198
\end{pmatrix}
\end{equation}
\end{document}